\documentclass[12pt]{article}

\usepackage{amsmath,amssymb,amsthm,mathrsfs,bbm}
\usepackage{latexsym,amscd,amsbsy,dsfont,amsfonts}
\usepackage{graphicx}
\usepackage[pdftex,colorlinks=true]{hyperref}
\newcommand{\hs}{\hspace{0.15mm}}

\newcommand{\ph}{\phantom{A}}

\usepackage{bm}
\usepackage{latexsym}
\usepackage[latin1]{inputenc}
\usepackage{amsfonts}
\usepackage{amssymb}
\usepackage{amsmath}
\usepackage{subfigure}

\numberwithin{equation}{section}

\begin{document}

\begin{flushright}
MCTP-14-29 \\
DCPT-14/35
\end{flushright}

\bigskip
\bigskip
\bigskip

\centerline{\Large \bf Schr\"odinger Holography for $z<2$}
\smallskip

\medskip
\bigskip
\bigskip
\centerline{\bf Tom\'as Andrade$^{a}$\footnote{tomas.andrade@durham.ac.uk}, Cynthia Keeler$^{b}$\footnote{keelerc@umich.edu}, Alex Peach$^{a}$\footnote{a.m.peach@durham.ac.uk}, {} and Simon F. Ross$^{a}$\footnote{s.f.ross@durham.ac.uk}}

\bigskip
\centerline{${}^{a}$Centre for Particle Theory, Department of Mathematical Sciences} 
\centerline{Durham University, South Road, Durham DH1 3LE, UK}
 \bigskip 
\centerline{${}^b$ Michigan Center for Theoretical Physics, Randall Laboratory of Physics}   
\centerline{The University of Michigan, Ann Arbor, MI 48109-1040, USA}
 
\bigskip
\bigskip

\begin{abstract}
We investigate holography for asymptotically Schr\"odinger spacetimes, using a frame formalism. Our dictionary is based on the anisotropic scaling symmetry. We consider $z <2$, where the holographic dictionary is cleaner; we make some comments on $z=2$. We propose a definition of asymptotically locally Schr\"odinger spacetime where the leading components of the frame fields provide suitable geometric boundary data.   We show that an asymptotic expansion exists for generic boundary data satisfying our boundary conditions for $z<2$.
\end{abstract}

\section{Introduction}

Holography for non-relativistic field theories has been actively studied for several years now. It has the potential to offer us tools to study a broader class of field theories holographically, which may include theories of interest for modelling condensed matter physics \cite{Son:2008ye,Balasubramanian:2008dm,Kachru:2008yh}. It also offers the possibility to deepen our understanding of holographic relations between field theories and gravity. The non-relativistic theories of interest are characterised by the existence of an anisotropic scaling symmetry which treats the time and space directions differently, $
t \to \lambda^z t, \quad x \to \lambda x$,
where $z$ is called the dynamical exponent. There are two main cases of interest, Schr\"odinger and Lifshitz. In the first case the theory has a Galilean boost symmetry; in the latter case there is no such symmetry, so the theory has a preferred rest frame.
As a result Schr\"odinger theories have a conserved particle number which is not present in the Lifshitz case.
The case $z=2$ is special for Schr\"odinger, in this case the theory has an additional special conformal symmetry. A holographic dual for theories with Schr\"odinger symmetry was proposed first \cite{Son:2008ye,Balasubramanian:2008dm}, but the Lifshitz case \cite{Kachru:2008yh} has been more fully explored, because of its greater simplicity and close resemblance to the well-understood AdS case.

For Lifshitz, the holographic dual has a metric
\begin{equation}
ds^2 = - \frac{dt^2}{r^{2z}} + \frac{d\vec{x}^2 +dr^2}{r^2},
\end{equation}
with $d_s$ spatial directions $\vec{x}$, which has an isometry $t \to \lambda^z t$, $x \to \lambda x$, $r \to \lambda r$ realizing the anisotropic scaling symmetry.
The bulk geometry has a single additional direction, $r$, related to energy scale in the dual field theory. Points for which $r \to 0$ are identified with the region in which the boundary
theory lives, although, due to the anisotropic scaling, there is no conformal boundary properly speaking.
Motivated by this, in \cite{Ross:2009ar} it was proposed that it is convenient to describe the geometry in terms of frame fields in constructing the holographic dictionary, and this dictionary was worked out in detail in \cite{Ross:2011gu}. This has been further developed in \cite{Baggio:2011cp,Mann:2011hg,Chemissany:2012du,Christensen:2013lma,Christensen:2013rfa}, and an alternative perspective based on a deformation of AdS for $z$ near one developed in \cite{Korovin:2013nha}.

For Schr\"odinger, the bulk metric is
\begin{equation} \label{schr}
ds^2 = -\frac{dt^2}{r^{2z}} + \frac{2 dt d\xi + d\vec{x}^2+ dr^2}{r^2}.
\end{equation}
Here we have once again chosen coordinates for which the boundary corresponds to $r \to 0$ in the sense explained above. The isometry $t \to \lambda^z t$, $x \to \lambda x$, $\xi \to \lambda^{2-z} \xi $ $r \to \lambda r$ realises the anisotropic scaling symmetry, and there are isometries $\vec{x} \to \vec{x} + \vec{v} t$, $\xi \to \xi - \vec v \cdot \vec x - \frac{1}{2} v^2 t$,  which realise the Galilean boost symmetry.  The presence of the additional null direction $\xi$ can be understood in field theory terms as arising from realizing the non-relativistic field theory with Galilean boosts as the light cone reduction of a Lorentz-invariant theory in one higher dimension \cite{Nishida:2007pj,Son:2008ye}. For $z=2$ this reduction is compatible with the anisotropic scaling symmetry. Thus, from the field theory point of view $\xi$ appears to play a kinematical role, as a useful device for realizing the symmetries of a Schr\"odinger theory in the more familiar framework of relativistic theories; momentum in the $\xi$ direction can be interpreted as a conserved particle number. Its role holographically has however remained somewhat unclear. The scaling symmetry acts non-trivially on $\xi$ for $z \neq 2$, so in this case it is only the higher-dimensional theory that is scale invariant, and it seems natural to assume that the holographic dictionary is formulated in terms of this higher-dimensional theory.

By a coordinate transformation $t \to \sigma t$, $\xi \to \sigma^{-1} \xi$, the metric \eqref{schr} can be rewritten as
\begin{equation}
ds^2 = -\frac{\sigma^2 dt^2}{r^{2z}} + \frac{2 dt d\xi + d\vec{x}^2+dr^2}{r^2}.
\end{equation}
and for small $\sigma$ the geometry outside of some neighbourhood of $r=0$ can be viewed as a deformation of AdS. This motivated the programme of \cite{Guica:2010sw,Costa:2010cn,Guica:2011ia,vanRees:2012cw}, which studies Schr\"odinger  holographically as the perturbation of a relativistic theory by an irrelevant vector operator, decomposing the linearised fluctuations of bulk fields in terms of sources and vevs of operators of given scaling dimension with respect to the relativistic scaling symmetry. This programme has had some success, but because the deforming operator is irrelevant, the understanding can only be perturbative in $\sigma$.

Our aim is instead to formulate a holographic dictionary based on identifying modes in the bulk with sources and vevs of operators of definite scaling dimension with respect to the anisotropic scaling symmetry, by applying the insights gained from the study of the Lifshitz case. Focusing on this non-relativistic perspective will allow us to treat the problem non-perturbatively. We formulate the dictionary in terms of frame fields. Such a formulation was attempted in \cite{Hartong:2013cba} for $z=2$, where an appropriate choice of frame fields and boundary conditions was identified, but difficulties were encountered in solving the equations of motion in an asymptotic expansion for general boundary conditions. One of our key insights is that it is easier to treat the case $z<2$, where the derivatives with respect to the boundary coordinates all come with powers of $r$, so dependence on these coordinates is negligible at leading order, and can be incorporated by adding appropriate subleading corrections to bulk fields.

The difference between $z <2$ and $z=2$ can be illustrated by considering a scalar field on a fixed Schrodinger background. The massive scalar wave equation $\Box \phi - m^2 \phi =0$ in the background \eqref{schr} is
\begin{equation}
r^2 \partial_r^2 \phi - (d_s + 1) r \partial_r \phi + r^2 (2 \partial_t \partial_\xi \phi + \partial_{\vec x}^2 \phi) + r^{4-2z} \partial_\xi^2 \phi - m^2 \phi = 0.
\end{equation}
For $z <2$, all of the derivatives along the boundary are suppressed at small $r$, appearing as $r^z \partial_t$, $r^{2-z} \partial_\xi$ and $r \partial_{\vec x}$, and the asymptotic radial falloff of the bulk solution is independent of the dependence on $t, \xi, \vec{x}$. Thus we can solve the equation in a power series in $r$, allowing the leading term in the series to have an arbitrary dependence on $t, \xi, \vec{x}$, and adding subleading corrections depending on derivatives of the leading term. For $z=2$ by contrast, the $\partial_\xi$ derivatives are not suppressed, and dependence on $\xi$ cannot be treated in this way.

Physically, this difference in the asymptotic expansion is due to a difference in the holographic dictionary. For $z<2$, $\phi$ (with the usual boundary conditions) is holographically dual to a local operator $O$ of dimension $\Delta= \frac{1}{2}(d_s+2) + \sqrt{ \frac{1}{4} (d_s+2)^2 + m^2}$ (with respect to the anisotropic scaling symmetry) which lives in a space parametrised by $t, \xi, \vec{x}$.\footnote{Also, for $z <2$ the anisotropic scaling symmetry acts non-trivially on the $\xi$ direction, so we have a scaling invariance only in this higher dimensional theory. If we restricted to the sector of a given momentum $k_\xi$, non-zero $k_\xi$ will break the scaling symmetry of expressions in the $t, \vec{x}$ space. So for instance the form of correlation functions in $t, \vec{x}$ is not constrained by the symmetry.} For $z=2$ by contrast, it is natural to decompose $\phi$ into Fourier modes, $\phi = \sum_{k_\xi} \phi_{k_\xi}(r, t, \vec{x})  e^{i k_\xi \xi}$, and identify each mode $\phi_{k_\xi}$ with a dual operator $O_{k_\xi}$ of dimension $\Delta = \frac{1}{2}(d_s+2) + \sqrt{ \frac{1}{4} (d_s+2)^2 + m^2 + k_\xi^2}$, living in a space parametrised by $t, \vec{x}$. In this $z=2$ case, the correlation functions of $O_{k_\xi}$ are constrained by the scaling symmetry.\footnote{ It is also interesting to note that once we restrict to a sector of fixed $k_\xi$, the scalar wave equation has a non-relativistic structure; the equation is first order in time derivatives.}

This distinction between $z<2$ Schr\"odinger and $z=2$ Schr\"odinger is analogous to the distinction between Lifshitz and the AdS$_2 \times \mathbb{R}^d$ geometry, which is the $z \to \infty$ limit of Lifshitz. For Lifshitz we think of the spatial directions as part of the space the field theory lives in, but for AdS$_2 \times \mathbb{R}^d$ the $\mathbb{R}^d$ directions are internal directions which are not affected by the scaling symmetry, and we think of the dual as a quantum mechanics living just in the time direction, with operators $O_{\vec{k}}$ labelled by the momentum in the $\mathbb{R}^d$ directions.

Thus, if we take the anisotropic scaling symmetry and the corresponding frame decomposition as the central elements in formulating the holographic dictionary, it is easier to work out the correspondence for $z<2$, where the dual theory naturally lives in all the $t, \xi, \vec{x}$ directions. The heart of our discussion will be a detailed treatment of $1 <z <2$ in the context of a massive vector theory, showing that it is possible to formulate the holographic dictionary in a familiar fashion, solving the equations of motion for given boundary data depending on $t, \xi, \vec x$ in an asymptotic expansion in powers of $r$, and constructing a well-behaved action by adding local boundary counterterms to the bulk action. As in the discussion of Lifshitz, the leading terms in the frame fields in the bulk will be interpreted as sources for the stress energy complex in the field theory.

Although we do not consider $z<1$ in this paper, we expect that our definitions and frame analysis can be equally well applied in that case.  For the other end of our range, $z=2$, our procedure will need modification. For $z=2$ we would want instead to formulate a dictionary in terms of a non-relativistic theory living in the $t, \vec{x}$ directions, where the different Fourier modes of the bulk fields are each thought of as corresponding to an operator in this field theory with $k_\xi$-dependent scaling dimension, as discussed above for a scalar field. The $\xi$ direction is at least asymptotically null, so we can't decompose the metric in a standard Kaluza-Klein reduction. However, in our holographic context it is more natural for us to think in terms of the frame fields, which are one-forms, which we can simply decompose into the component along $d\xi$ and the components along the remaining boundary directions. In this $z=2$ case, the zero-modes in the leading terms in the frame fields in the bulk will be interpreted as sources for the stress energy complex in the non-relativistic field theory living in the $t, \vec{x}$ directions. In addition, for $z=2$ there are potential logarithmic terms in the asymptotic expansion which need to be treated carefully. We therefore leave a detailed study of $z=2$ for future work.

We start in the next section by reviewing the Schr\"odinger solution in a little more detail, introducing the massive vector theory we will work in for the remainder of this paper (although it should be easy to extend these ideas to alternative realizations of Schr\"odinger such as topologically massive gravity). We introduce our frame decomposition of the metric following \cite{Hartong:2013cba} and discuss how the frame rotation symmetry can be partially fixed by relating the frame fields to the massive vector. We then define our asymptotically locally Schr\"odinger boundary conditions in terms of these frame fields. In section \ref{stress}, we review the structure of the stress energy complex for non-relativistic theories, and discuss the description in the higher-dimensional theory including the $\xi$ direction. In section \ref{lin} we give a linearised analysis around the Schr\"odinger solution for $z <2$, and identify the linearised modes with sources and vevs for the stress energy complex and matter operator. In section \ref{asymp}, we discuss the asymptotic expansion for $z<2$, and show that a solution can be obtained in an expansion in powers of $r$,\footnote{In our analysis, this is traded for an expansion in eigenvalues of a suitable dilatation operator, but the existence of a dilatation expansion implies the existence of an expansion in powers of $r$, since each term in the dilatation expansion has an expansion in positive powers of $r$.}  and that all divergences in the action can be eliminated by adding boundary counterterms which are local functions of the boundary data.

\section{Asymptotically locally Schr\"odinger boundary conditions}
\label{bc}

We consider the metric \eqref{schr} as a solution of the theory with a massive vector introduced in \cite{Son:2008ye}. The action is
\begin{equation} \label{action}
S =-\frac{1}{16 \pi G} \int d^{d_s + 3} x \sqrt{-g} \left(R -2 \Lambda - \frac{1}{4} F_{\mu\nu} F^{\mu\nu} - \frac{1}{2} m^2 A_\mu A^\mu \right) - \frac{1}{8 \pi G} \int d^{d_s +2} \xi \sqrt{-\gamma} K,
\end{equation}
where $\gamma$ is the induced metric on the boundary and $K$ is the trace of the extrinsic curvature, with
\begin{equation}
	m^2 = z(z+d_s), \quad \Lambda = - \frac{(d_s+2)(d_s+1)}{2}.
\end{equation}
The equations of motion that follow are
\begin{equation}\label{eins eqs}
	R_{\mu \nu} - \frac{1}{2} R g_{\mu \nu} + \Lambda g_{\mu \nu}
	= \frac{1}{2} \left(  F^\alpha \hs _\mu F_{\alpha \nu} - \frac{1}{4} F^2 g_{\mu \nu}   \right)
	+ \frac{m^2}{2}\left(  A_\mu A_\nu - \frac{1}{2} A^2 g_{\mu \nu}  \right)
\end{equation}
\begin{equation}\label{maxw eq}
	\nabla_\mu F^{\mu \nu} = m^2 A^\nu
\end{equation}
The metric \eqref{schr} is a solution of \eqref{eins eqs}, \eqref{maxw eq} supported by the matter field
\begin{equation}\label{A sch}
	A = \alpha  r^{-z} dt , \quad \alpha  = \sqrt{\frac{2(z-1)}{z}}.
\end{equation}
The massive vector field $A_\mu$ physically singles out the $t$ direction as special.

We want to define a class of asymptotically locally Schr\"odinger spacetimes which asymptotically approach \eqref{schr} locally as $r \to 0$. Inspired by the analysis in the Lifshitz case, it is natural to do so by introducing an appropriate set of frame fields. We will adopt the frame decomposition proposed in  \cite{Hartong:2013cba} ,
\begin{equation} \label{a metric}
ds^2 = g_{AB} e^A e^B =  - e^+ e^+ + 2 e^+ e^- + e^I e^I + e^r e^r,
\end{equation}
for some frame fields $e^A$, $A= +,-,I,r$. We will always adopt the radial gauge choice $e^r = r^{-1} dr$.  In the background \eqref{schr} $e^+ = r^{-z} dt$, $e^- = r^{z-2} d\xi$, $e^I = r^{-1} dx^i$, so each of the frame fields has a well-defined scaling with $r$ at large $r$.\footnote{Note that for this flat background, the frame index $I$ and the coordinate index $i$ are equivalent.} Note the main novelty compared to more familiar cases is that to achieve this simple form for the individual frame fields, we take the frame metric $g_{AB}$ to have off-diagonal components.

The decomposition of the metric does not fix the choice of frame fields uniquely; it is invariant under transformations which preserve the metric $g_{AB}$, so we have the freedom to redefine the $e^A$ infinitesimally by
\begin{equation} \label{boost}
e^+ \to e^+ + \alpha^I e^I, \quad e^- \to e^- + \beta^I e^I, \quad e^I \to e^I - \beta^I e^+ + \alpha^I (e^+ - e^-)
\end{equation}
and
\begin{equation} \label{gamma}
e^+ \to e^+ + \gamma e^+, \quad e^- \to e^- + \gamma (e^+ - e^-).
\end{equation}
The decomposition is also invariant under rotations among the spatial frame fields $e^I$. We could leave this symmetry unfixed in the spirit of the treatment of the Lifshitz case in \cite{Christensen:2013lma,Christensen:2013rfa}, but we prefer to relate the distinguished frame fields to physical quantities, fixing this symmetry as much as possible. This will simplify the task of identifying the sources for the operators in the stress energy complex.

In our massive vector theory, the symmetries (\ref{boost},\ref{gamma}) will be restricted by assuming a form for the massive vector field. We can first restrict \eqref{boost} by assuming $A$ has no $e^I$ component, so
\begin{equation}
A = A_+ e^+ + A_- e^- + r A_r e^r.
\end{equation}
The transformations which preserve this are those with $A_+ \alpha^I + A_- \beta^I =0$, together with the rotations of the spatial frame fields. The action of
\eqref{gamma} is
\begin{equation}
A_+ \to A_+ + \gamma (A_+ + A_-), \quad A_- \to A_- - \gamma A_-.
\end{equation}
Since the frame field $e^+$ is a null vector, it doesn't have a fixed length. The symmetry \eqref{gamma} rescales it; we can therefore use this flexibility to fix the value of the projection of $A$ along $e^+$.  A convenient choice is to set $A_+ = \alpha$, its background value.

Thus we choose
\begin{equation} \label{a vector}
A = \alpha (e^+ + \psi e^-  + s_r e^r),
\end{equation}
where $\alpha$ is the constant background value in \eqref{A sch}, and $\psi$ is the single scalar degee of freedom in the boundary conditions for the matter field, and we've taken an overall factor of $\alpha$ out for convenience. We will find that the operator dual to $\psi$ is irrelevant, so we always set the source part to zero. 

Given any solution of the massive vector theory, we can write the metric and vector field as in (\ref{a metric},\ref{a vector}). The physical degrees of freedom are then the frame fields $e^A$ and the scalar $\psi$. As in Lifshitz, a part of the degrees of freedom in the massive vector field has been assigned to the frame fields, to make physical some of the components that would have been pure gauge. Unlike in Lifshitz, this does not make all of the components of $e^+$, $e^-$ physical. The remaining gauge symmetry is
\begin{equation} \label{res}
e^+ \to e^+ - \psi \beta^I e^I, \quad e^- \to e^- + \beta^I e^I, \quad e^I \to e^I - \beta^I e^+ - \psi \beta^I (e^+ - e^-),
\end{equation}
together with the rotations of the spatial frame fields $e^I$.

We then say that a spacetime is {\it asymptotically locally Schr\"odinger} if the metric and massive vector can be written as in (\ref{a metric},\ref{a vector}) with
\begin{equation} \label{ALS}
e^+ = r^{-z} \hat e^+, \quad e^- = r^{z-2} \hat e^-, \quad e^I = r^{-1} \hat e^I,
\end{equation}
and the scalar $\psi = r^{\Delta_-} \hat \psi$ for some exponent $\Delta_-$,%
\footnote{This leading asymptotic falloff of the scalar will be determined later by the linearised analysis, where for $d_s=2$ we find that $\Delta_- = 2-2z$, corresponding to a scalar operator of dimension $2z+2$ in the dual field theory.}
 where the fields $\hat e^A, \hat \psi$ are arbitrary functions of $t, \xi, \vec{x}, r$ with finite limits as $r \to 0$. The boundary limits of the $\hat e^A$  (which with characteristic abuse of notation we will sometimes refer to simply as $\hat e^A$) define the boundary geometry for our asymptotically locally Schr\"odinger spacetime (while the scalar $\hat \psi$ is the source for a scalar operator in the dual field theory).

\section{Stress energy complex and dimensional reduction}
\label{stress}

We want to view this data as describing the geometry our field theory lives in, so it should provide sources for the stress energy complex. Let us therefore review the structure of this in a non-relativistic theory. Any non-relativistic theory, Lifshitz or Schr\"odinger, will have an energy density $\mathcal E$ and an energy flux $\mathcal E^i$, satisfying the conservation equation (in a flat boundary space) 
\begin{equation} \label{nre}
\partial_t \mathcal E + \partial_i \mathcal E^i =0,
\end{equation}
along with a momentum density $\mathcal P_i$ and a spatial stress tensor $\Pi_{ij}$ satisfying the conservation equation
\begin{equation} \label{nrp}
\partial_t \mathcal P_i + \partial_j \Pi_i^j = 0.
\end{equation}
The Schrodinger theory additionally has a conserved particle number, so there is a particle number density $\rho$ and a particle number flux $\rho^i$ satisfying
\begin{equation} \label{nrrho}
\partial_t \rho + \partial_i  \rho^i = 0.
\end{equation}
The scale invariance implies $z \mathcal E + \Pi_i^i + (2-z) \rho =0$; the additional term for $z \neq 2$ is associated with the breaking of the scaling symmetry by non-zero particle number. $\mathcal E$ has dimension $z+d_s$, which implies $\mathcal E^i$ has dimension $2z +d_s-1$, and $\mathcal P_i$ has dimension $1+d_s$, which implies $\Pi_{ij}$ has dimension $z+d_s$. The particle number has dimension $2-z$, so its density $\rho$ has dimension $2-z+d_s$, so $\rho^i$ has dimension $1+d_s$. In fact, in a non-relativistic theory $\rho^i = \mathcal P_i = \rho v^i$, where $v^i$ is the local velocity of the particles, so these are not independent operators.

In the Lifshitz story the stress energy complex was realised directly in the holographic dual, but in Schrodinger the non-relativistic field theory is constructed as the reduction of a one higher dimensional field theory over a null circle labelled by the coordinate $\xi$. For $z <2$, it is the higher-dimensional quantities that we expect to appear in our holographic dictionary. In \cite{Rangamani:2008gi}, non-relativistic quantities were obtained by dimensional reduction from the stress tensor of a relativistic theory. In the present paper, we work in a frame formalism adapted to the anisotropic scaling symmetry, so the description in the higher-dimensional theory is still not relativistic; in particular different components have different scaling dimensions even in the higher-dimensional description.

In the higher-dimensional theory for $z <2$, we expect to have an energy current whose sources are in the frame field $\hat e^+$, a $\xi$-momentum current which is physically identified with particle number whose sources are in $\hat e^-$, and spatial momentum currents whose sources are in $\hat e^I$.  The energy current consists of an energy density $E$, an energy flux $E^i$ in the spatial directions, and an energy flux $E^\xi$ in the null direction. The conservation equation is
\begin{equation} \label{he}
\partial_t E + \partial_i E^i + \partial_\xi E^\xi=0.
\end{equation}
The relation between these $d_s+2$ dimensional operators and the above $d_s+1$ dimensional theory is that the densities in the $d_s+1$ dimensional theory are the integral of the higher-dimensional densities over the $\xi$ circle, so $\mathcal E = \oint d\xi E$ etc. Thus $E$ has dimension $d_s+2$, so that integrating over $d\xi$ (which has dimension $z-2$) gives $\mathcal E$ dimension $z+d_s$. This can also be understood directly in the higher-dimensional theory;  the densities in this theory are per unit volume in $\vec x$ and $\xi$. The volume element $d\xi \, d^{d_s} x$ has dimension $z-2-d_s$, so $E$ has dimension $d_s+2$ so the total energy obtained by integrating over the volume element has dimension $z$. The spacetime volume element in $d_s+2$ dimensions has length dimension $d_s+2$ with respect to the anisotropic scaling,  so this is the dimension of a marginal operator. The conservation equation implies $E^i$ has dimension $z+d_s+1$ and $E^\xi$ has dimension $2z+d_s$, as $\partial_\xi$ has dimension $2-z$.

The spatial momentum currents similarly consist of the spatial momentum density $P_i$, a stress tensor $T_{ij}$ in the spatial directions, and a stress $T_i^\xi$ in the $\xi$ direction, satisfying the conservation equation
\begin{equation} \label{hp}
\partial_t P_i + \partial_j T_i^j + \partial_\xi T_i^\xi= 0.
\end{equation}
$P_i$ has dimension $3-z+d_s$, so that the total momentum has dimension 1, and the integral over $\xi$ gives $\mathcal P_i = \oint d\xi P_i$ dimension $d_s+1$ as expected in a non-relativistic theory. The conservation equation then implies that $T_{ij}$ has dimension $d_s+2$ and $T_i^\xi$ has dimension $z+d_s+1$.

Finally, the $\xi$-momentum current consists of the momentum density $P_\xi$ in the $\xi$ direction, which will be identified with particle number density. This density comes with a particle number flux $P_\xi^i$ in the spatial directions and  $P_\xi^\xi$ in the $\xi$ direction, satisfying
\begin{equation} \label{hpxi}
\partial_t P_\xi + \partial_j P_\xi^j + \partial_\xi P_\xi^\xi= 0.
\end{equation}
$P_\xi$ has dimension $4-2z+d_s$, implying $P_\xi^i$ has dimension $3-z+d_s$ and $P_\xi^\xi$ has dimension $d_s+2$.As noted earlier, $P_\xi^i = P_i$, and $T_{ij}$ is a symmetric tensor. The Ward identity from the scaling symmetry is $z E + T_i^i + (2-z) P_\xi^\xi = 0$.\footnote{The Ward identities (\ref{nre},\ref{nrp},\ref{nrrho}) and $z \mathcal E + \Pi_i^i + (2-z) \rho =0$ are obtained by taking the above identities and integrating over the $\xi$ circle.}

Apart from these Ward identities the components of the stress complex are independent; note in particular that $T_i^\xi$ and $P_\xi^i$ have different dimensions, so the stresses in the spatial and $\xi$ directions cannot be combined into a symmetric tensor. Note that  $E^\xi$, $E^i$ and $T_i^\xi$ are irrelevant operators.

For $z <2$, our holographic dictionary will naturally be formulated in terms of this $d_s +2$ dimensional field theory, and the frame fields $\hat e^A$ provide sources for the corresponding currents, which can be arbitrary functions of $t, \xi, \vec{x}$. We can view these currents as the components of the non-symmetric tensor
\begin{equation} \label{TalphaB}
T^{\alpha}_{\ \ B} = \frac{1}{\sqrt{-\gamma}}\frac{\delta}{\delta e_{\alpha}^{B}}S.
\end{equation}
The residual gauge symmetry \eqref{res} corresponds to the fact that there are not independent physical sources for $P_i$, $P_\xi^i$, while the symmetry under rotations of the $\hat e^I$ corresponds to $T_{ij}$ being a symmetric tensor.

As in the Lifshitz case, there are irrelevant operators in the stress energy complex, and we would expect to need to set their sources to zero. For generic sources, there is no diffeomorphism-invariant part in the source for $T_i^\xi$, as we can always make a $\xi$-dependent redefinition of the $x^i$ coordinates to set the $d\xi$ components in $e^I$ to zero.\footnote{The zero-mode of the source of $T_i^\xi$ along the $\xi$ direction is diffeomorphism-invariant, so in the discussion of $z=2$ we will have to explicitly set this to zero.}  Therefore the only diffeomorphism-invariant sources for irrelevant operators are in $e^+$, and we can set these to zero by adopting the irrotational condition
\begin{equation}
\hat e^+ \wedge d \hat e^+= 0.
\end{equation}
As in Lifshitz, this can be viewed as a condition that the boundary geometry defined by the $\hat e^A$ admits a foliation by surfaces of absolute time, as is appropriate for a non-relativistic theory. As in Lifsihtz we will find that there is a range of values of $z$ for which solutions exist in an asymptotic expansion  even if we do not impose this condition. Since the energy flux $E^i$ is irrelevant for all $z >1$, one might expect that we would always need to set its source to zero. But the diffeomorphism symmetry implies that only derivatives of this source actually appear, so there is a range of values for which the asymptotic expansion exists even in the presence of this source, as in Lifshitz.  Here the relevant range is $z < 3/2$. 

For $z=2$, the story is different. We argued in the introduction that because the anisotropic scaling symmetry doesn't act on $\xi$ and the asymptotic falloffs of bulk modes of different $k_\xi$ are different,  the appropriate holographic dictionary is now in terms of a theory that lives in $d_s +1$ dimensions, with modes of different $k_\xi$ identified with distinct operators in this theory, whose scaling dimensions may be $k_\xi$ dependent. Thus, to identify the boundary data $\hat e^A, \psi$ as sources for the dual operators, we should expand them in Fourier modes in $\xi$. For the frame fields, we should also decompose them as
\begin{equation}
\hat e^A = \hat e^A_a dx^a + \hat e^A_\xi d\xi,
\end{equation}
where $a$ runs over $t, x^i$. For the zero modes, where $\hat e^A$ is independent of $\xi$, this decomposition is the analogue in our frame language of the Kaluza-Klein decomposition of the metric and massive vector field. With respect to the $\xi$-independent diffeomorphisms acting in the lower-dimensional boundary  coordinates, $e^A_a$ will transform as a one-form and $e^A_\xi$ will transform as a scalar.

As noted above, the operators in the stress complex in the $d_s +1$  dimensional non-relativistic theory are obtained by integrating the higher-dimensional densities over the $\xi$ circle, $\mathcal E = \oint d\xi E$ etc. That is, they are the zero modes of the higher-dimensional fields along this circle direction. The sources for these operators are thus the $\xi$-independent part of the sources $\hat e^A_a$. The conservation equations (\ref{nre} - \ref{nrrho}) are obtained by integrating (\ref{he} - \ref{hpxi}) over the $\xi$ circle; the last terms in the latter equations will drop out on doing the integral as they are a total derivative. Thus, for $z=2$, we could obtain correlation functions of the non-relativistic stress energy complex just by considering appropriate $\xi$-independent sources  $\hat e^A_a$. We can also consider $\xi$-independent sources $\hat e^A_\xi$, which are interpreted for $z=2$ as providing sources for some particular scalar operators.\footnote{The situation is similar to the Lifshitz theories obtained by dimensional reduction in \cite{Chemissany:2012du}.}

\section{Linearised analysis for $z <2$}
\label{lin}

We now turn to a linearised analysis of the equations of motion (\ref{eins eqs},\ref{maxw eq}) for $z <2$. We will see that this analysis confirms that the limits as $r \to 0$ of the rescaled frame fields $\hat e^A_\alpha$ can be interpreted as the sources corresponding to the stress energy complex $T^\alpha_{\ \ B}$, in that the modes canonically conjugate to the sources in the symplectic flux satisfy the expected Ward identities as a consequence of the linearised equations in the bulk. We will identify $\psi$ as the bulk dual of an operator of dimension $2z+2$ when $d_s=2$. We will see that the equations can be solved in a power series in $r$ in the asymptotic region, where the subleading terms are determined locally in terms of the sources.

We will consider the case $d_s=2$, which is physically the most interesting (the results for other values of $d_s$ will be similar in structure) and $d_s=0$, which is a special case and was previously analysed in \cite{Costa:2010cn}, so discussing this case will be useful for comparison purposes.

The linearised version of our frame fields is
\begin{eqnarray}
\hat e^+ &=& (1 + \delta \hat e^+_t) dt +  \delta \hat e^+_\xi d\xi +  \delta \hat e^+_i dx^i, \\
\hat e^- &=& (1 + \delta \hat e^-_\xi) d\xi + \delta \hat e^-_t dt +  \delta \hat e^-_i dx^i, \\
\hat e^I &=&  (\delta^I_j + \delta \hat e^I_j) dx^j + \delta \hat e^I_t dt +  \delta \hat e^I_\xi d\xi.
\end{eqnarray}
The linearised fields are then $\delta \hat e^A_\alpha$ and the $\psi$, $s_r$ in \eqref{a vector}. The constant modes in $\delta \hat e^A_\alpha$ are assumed to represent sources for the corresponding components of $T^\alpha_{\ A}$.

The linearised version of the residual gauge symmetry \eqref{res} is $ \delta \hat e^-_i \to \delta \hat e^-_i + \hat \beta^i $, $\delta \hat e^I_t \to \delta \hat e^I_t - \hat \beta^i $ (where $\beta^I = r^{z-1} \hat \beta^i$). This implies that the sources for $T^+_{\ \ I} = P_i$ and $T^I_{\ \ \xi} = P_\xi^i$ are not independent, as expected. The rotation symmetry of the $e^I$ also implies that only the symmetric part of $\delta e^I_j$ provide independent sources. The equations of motion are easier to discuss in the metric language, so we will resolve this gauge symmetry by passing back from the frame fields to the metric and vector for this linearised analysis.

In the metric language, the linearised perturbations are $h_{\mu\nu}$, $a_\mu$.
The linearised equations in the metric language are as in \cite{Ross:2009ar}\footnote{Note that $h_{\mu\nu}$ denotes
  the perturbation of the metric, and indices are raised and lowered
  with the background metric, so $h^{\mu\nu}$ is the perturbation of
  the metric with the indices raised, not the perturbation of the
  inverse metric.}
\begin{equation}
\nabla_\mu f^{\mu\nu} - \nabla_\mu (h^{\mu\lambda} F_\lambda^{\ \nu})
- \nabla_\mu h^{\beta \nu} F^\mu_{\ \beta}
+ \frac{1}{2} \nabla_\lambda h F^{\lambda \nu} = m^2 a^\nu
\end{equation}
and
\begin{eqnarray}
R_{\mu\nu}^{(1)} &=& \frac{2}{d-2} \Lambda h_{\mu\nu} + \frac{1}{2} f_{\mu\lambda} F_\nu^{\ \lambda} +
\frac{1}{2} f_{\nu\lambda} F_\mu^{\ \lambda} - \frac{1}{2}
F_{\mu\lambda} F_{\nu\sigma} h^{\lambda \sigma} -\frac{1}{2(d-2)}
f_{\lambda \rho} F^{\lambda \rho} g_{\mu\nu}   \nonumber \\ &&+ \frac{1}{2(d-2)} F_{\lambda
  \rho} F_\sigma^{\ \rho} h^{\lambda \sigma} g_{\mu\nu}- \frac{1}{4(d-2)}
F_{\lambda \rho} F^{\lambda \rho} h_{\mu \nu} + \frac{1}{2} m^2
a_\mu A_\nu + \frac{1}{2} m^2 a_\nu A_\mu,
\end{eqnarray}
where $d=d_s+3$ is the dimension of the spacetime, $f_{\mu\nu} = \partial_\mu a_\nu - \partial_\nu a_\mu$ and
\begin{equation}
R_{\mu\nu}^{(1)} = \frac{1}{2} g^{\lambda \sigma} [ \nabla_\lambda
  \nabla_\mu h_{\nu \sigma} + \nabla_\lambda \nabla_\nu h_{\mu\sigma} -
  \nabla_\mu \nabla_\nu h_{\lambda \sigma} - \nabla_\lambda
  \nabla_\sigma h_{\mu\nu} ].
\end{equation}

It is convenient to write
\begin{equation}
	h_{tt} = r^{-2z} H_{tt},  \qquad h_{t \xi} = r^{-2} H_{t \xi},  \qquad h_{\xi \xi} = r^{2(z-2)} H_{\xi \xi},
\end{equation}
\begin{equation}
	h_{ti} =  r^{-(z+1)} H_{ti} \qquad h_{\xi i} = r^{z-3} H_{\xi i}, \qquad h_{ij} = r^{-2} H_{ij},
\end{equation}
\begin{equation}
	a_r = \alpha r^{-1} s_r  \qquad a_t = \alpha r^{-z} s_t  \qquad a_\xi = \alpha r^{z-2}  s_\xi  \qquad  a_{i} = \alpha r^{-1} s_{i}.
\end{equation}
Then, a given linearised mode will contribute at the same order in $r$ in all the different fields, and the power of $r$ will correspond to the scaling dimension of the mode. The $s_r$ here is the same as in \eqref{a vector}, and the other fields are related to the linearised frame fields by
\begin{equation} \label{lin map 1}
H_{tt} = -2 \delta \hat e^+_t + 2 r^{2z-2} \delta \hat e^-_t, \quad H_{t\xi} = - r^{2-2z} \delta \hat e^+_\xi  + \delta \hat e^-_\xi + \delta \hat e^+_t, \quad H_{\xi\xi} = 2 r^{2-2z} \delta \hat e^+_\xi,
\end{equation}
\begin{equation}\label{lin map 2}
H_{ti} = - r^{1-z} \delta \hat e^+_i + r^{z-1} \delta \hat e^-_i + r^{z-1} \delta \hat e^I_t, \quad H_{\xi i} = r^{1-z} \delta \hat e^+_i + r^{1-z} \delta \hat e^I_\xi, \quad H_{ij} =  \delta \hat e^I_j + \delta \hat e^J_i,
\end{equation}
\begin{equation} \label{lin map 3}
s_t = \delta \hat e^+_t, \quad s_\xi = r^{2-2z} \delta \hat e^+_\xi + \psi, \quad s_i = r^{1-z} \delta \hat e^+_i.
\end{equation}
Note that in the expansion about a flat background the $I$ and $i$ indices are equivalent at leading order, so in these equations, $\delta \hat e^I_\alpha$ should be understood as $\delta \hat e^I_\alpha \delta_{Ii}$. 

\subsection{Linearised solutions for $d_s=2$}

Let us now study the equations for $d_s=2$. Our interest is in understanding the identification of the solutions of the linearised equations with sources and vevs for the dual operators. We identify the sources with the leading constant parts of the linearised frame fields $\delta \hat e^A$, which appear in the linearised fields in the frame language as set out in (\ref{lin map 1} - \ref{lin map 3}).  Since we have not yet carried out a holographic renormalization procedure, the vevs will also have divergent contributions from the source modes, but we are interested in identifying the relation between the bulk solutions which are not locally determined by the sources and the finite part of the vevs. In many cases, we can identify the mode corresponding to the vev by its conformal dimension alone, but in general we follow \cite{Papadimitriou:2010as} and identify the vev as the linearised solution which is canonically conjugate to the source with respect to the symplectic inner product defined by calculating the symplectic flux.

Since dependence on the boundary directions introduces only subleading terms for $z <2$, we can first understand this identification by considering constant modes, which are independent of the boundary directions. We then discuss briefly the linearised equations for non-constant modes and check that the solutions we are identifying with the vevs do indeed satisfy appropriate Ward identities as a result of the asymptotic equations of motion in the non-constant cases.

When the fields are independent of spatial coordinates $x^i$, the rotation symmetry in these directions will be unbroken, so we can decompose the linearised fields into a tensor, vector and scalar part with respect to this linearised symmetry. Below we will treat these tensor, vector and scalar modes first, initially for constant modes and then including dependence on $t, \xi$. To make this decomposition we should further decompose $H_{ij}$ into a trace and a trace free part, $H_{ij} = k \delta_{ij} + \bar H_{ij}$, where $\bar H_i^i = 0$.  The tensor mode is $\bar H_{ij}$. The vector modes are $H_{ti}$, $H_{\xi i}$ and $s_i$. The scalar modes are $H_{tt}$, $H_{t\xi}$, $H_{\xi\xi}$, $k$, $s_t$, $s_\xi$ and $s_r$ (which is determined algebraically in terms of the other modes). We will always assume the $t,\xi$ dependence is harmonic, $e^{i \omega t + i k_\xi \xi}$, so in writing equations we will make the replacements $\partial_t \to i \omega$, $\partial_\xi \to i k_\xi$.

When we include dependence on the $x^i$, there is a different decomposition, which splits the modes into scalars (which now include scalar-derived vectors and tensors) and vectors (including vector-derived tensors). We set up the equations for this general case in section \ref{spatial}, and comment on the Ward identities.

\subsubsection{Tensor modes}

The tensor equation of motion is
\begin{equation} \label{tensoreq}
r^2 \bar H_{ij}'' - 3r \bar H_{ij}' - (  k_\xi ^2 r^{2(2-z)} + 2 k_\xi \omega r^2  ) \bar H_{ij} = 0.
\end{equation}
The solution for $\omega = k_\xi =0$ is
\begin{equation}
\bar H_{ij} = \bar H_{ij}^{(0)} + \bar H_{ij}^{(4)} r^4,
\end{equation}
corresponding to the source and vev for the trace free part of the spatial stress tensor $T_{ij}$. For general $(k_\xi,\omega)$, we will have an infinite series of subleading corrections which involve boundary derivatives of these leading terms. As the equation of motion only involves the combinations $k_\xi \omega$ and $k_\xi^2$, the  solution can be written as
\begin{equation} \label{tensor exp}
\bar H_{ij} = \sum_{m,n \geq 0} a_{ij(m,n)}  (k_\xi \omega r^2)^{2 m} (k_\xi r^{2-z})^{2n}
	+ k_\xi^2 \omega^2 r^4 \log r^2 \sum_{m,n \geq 0} b_{ij(m,n)}  (k_\xi \omega r^2)^{2 m} (k_\xi r^{2-z})^{2n} .
\end{equation}
We can take $a_{ij(0,0)} = \bar{H}_{ij}^{(0)}$ and $a_{ij(2,0)} = \bar H_{ij}^{(4)}$ as the independent coefficients. The expansion includes log terms because a subleading term determined by $ \bar{H}_{ij}^{(0)}$  and the independent term $\bar H_{ij}^{(4)}$ occur  at the same power of $r$. The subleading terms in the expansion are all determined in terms of $\bar{H}_{ij}^{(0)}$ and $\bar H_{ij}^{(4)}$ by solving \eqref{tensoreq} in a power series in $k_\xi$, $\omega$. The explicit factors of $k_\xi$, $\omega$ in \eqref{tensor exp} imply that there will be no factors of  $k_\xi$, $\omega$ in the equations for the $a_{ij(m,n)}$, $b_{ij(m,n)}$, so the subleading terms are determined locally in the boundary directions. They are solutions of ODEs in the radial direction.

\subsubsection{Vector modes}

The vector equations of motion are
\begin{align}
\label{vi xi kxi omega}
	r^2 s_i'' - 3 r s_i' - [ (z-1)(z+3) + k_\xi^2 r^{4-2z}  + 2 k_\xi \omega r^2 ]  s_i + z r H_{\xi i}' + z(z-1) H_{\xi i} &= 0, \\	
\label{Hti xi kxi omega}
	k_\xi [  r ( H_{\xi i}' + H_{t i}' ) + (z-1) ( H_{\xi i} - H_{t i} - 2  s_i)   ] + \omega r^{2z-2} [ r H_{\xi i}' + (z-1) H_{\xi i}  ] &= 0 , \\
\label{Hxii xi kxi omega}
	r^2 H_{\xi i}'' + (2z - 5) r H_{\xi i}' + [(z-1)(z-5)  - r^2 k_\xi \omega  ] H_{\xi i} + k_\xi^2 r^{4-2z} H_{ti} & = 0,
\end{align}
and
\begin{align}
\nonumber
	r^2 H_{ti}'' & -  r (2z+1) H_{ti}' + [ (z-1)(z+3) - k_\xi^2 r ^{4 - 2z} - r^2 k_\xi \omega] H_{ti} \\
\label{2nd order Hti kxi omega}
	& + 2(z-1) [ (z+3) s_i - (z-1) H_{\xi i} - r (s_i + H_{\xi i})' ] + (r^2 k_\xi \omega + r^{2z} \omega^2) H_{\xi i}= 0 .
\end{align}
For $k_\xi = \omega = 0$, \eqref{Hti xi kxi omega} is trivially satisfied, and we solve (\ref{vi xi kxi omega},\ref{Hxii xi kxi omega},\ref{2nd order Hti kxi omega}). For general $k_\xi$, $\omega$, we solve  (\ref{vi xi kxi omega},\ref{Hti xi kxi omega},\ref{Hxii xi kxi omega}), which imply (\ref{2nd order Hti kxi omega}).

For $k_\xi = \omega = 0$, the solution for the vector modes can be written as
\begin{align}
	H_{\xi i} &= (s_i^{(-)} + H_{\xi i}^{(-)}) r^{1-z}  +  H_{\xi i}^{(+)}   r^{5-z},   \\
	H_{t i} &= - s_i^{(-)} r^{1-z} + H_{t i}^{(-)} r^{z-1}  +  H_{t i}^{(+)} r^{z+3}   +  \frac{(z-4)}{2(3-z)}  H_{\xi i}^{(+)}  r^{5 - z}, \\
	s_i & = s_i^{(-)} r^{1-z}  + \frac{z}{2(z-1)} H_{\xi i}^{(+)} r^{5-z}  + s_i^{(+)} r^{z+3}.
\end{align}
We have chosen to define and normalise the independent so that the solutions with a $(-)$ superscript correspond to the sources, coming from the constant modes in the frame fields. From \eqref{lin map 3}, we see that $s_i^{(-)}$ corresponds to the constant part in $\delta \hat e^+_i$, the source term for the energy flux $E^i$. From \eqref{lin map 2}, $ H_{\xi i}^{(-)}$ is then the constant part of $\delta \hat e^I_\xi$, the source term for the stress $T^\xi_i$, and $H_{t i}^{(-)}$ is the source term for the momentum density $P_i$. The modes with a $(+)$ superscript should then correspond to the vevs of these operators. By dimensions alone we see that $\langle P_i \rangle \sim H_{\xi i}^{(+)}$. The vevs $\langle E^i \rangle$ and $\langle T^\xi_i \rangle$ should be related to $H_{t i}^{(+)}$ and $s_i^{(+)}$.

We can work out the identification by computing the symplectic flux at the boundary $r=0$, and identifying the modes canonically conjugate to the sources with the vevs, following \cite{Papadimitriou:2010as}.
Generically, the symplectic flux will have divergent contributions involving just the source modes, corresponding to the divergences in the vevs which need to be removed by holographic renormalization, but we focus on constant perturbations for which the result is finite, enabling us to relate the $(+)$ modes to the finite part of the vevs.
The appropriate symplectic current for the Einstein-massive vector theory we are considering was worked out in  \cite{Andrade:2013wsa}. It involves combining the usual gravitational symplectic current $j^\mu_g$ with an additional component $j^\mu_a$,
\begin{equation}
	j^\mu = j^\mu_g + j^\mu_a.
\end{equation}
These are respectively given by
\begin{equation}
j^\mu_g = P^{\mu\nu\alpha\beta\gamma\delta} (h_{2 \alpha \beta}^* \nabla_\nu h_{1 \gamma \delta} - h_{1 \alpha \beta} \nabla_\nu h_{2 \gamma \delta}^*),
\end{equation}
\begin{equation}
j^\mu_a = a_{2 \nu}^* (f_1^{\mu \nu} - h_1^{\mu\lambda} F_\lambda^{\ \nu} - h_1^{\beta \nu} F^\mu_{\ \beta} + \frac{1}{2} h_1 F^{\mu \nu}) - (1 \leftrightarrow 2),
\end{equation}
where
\begin{equation}
P^{\mu\nu\alpha\beta\gamma\delta} = \frac{1}{2} (g^{\mu\nu} g^{\gamma (\alpha} g^{\beta) \delta} + g^{\mu (\gamma} g^{\delta) \nu} g^{\alpha \beta} + g^{\mu (\alpha} g^{\beta) \nu} g^{\gamma \delta} - g^{\mu\nu} g^{\alpha \beta} g^{\gamma \delta} - g^{\mu (\gamma} g^{\delta) (\alpha} g^{\beta) \nu} - g^{\mu (\alpha} g^{\beta) (\gamma} g^{\delta) \nu}),
\end{equation}
indices in parentheses are symmetrized, and $*$ indicates complex conjugation.

Given the current found from two linearised solutions, the symplectic flux through the boundary, ${\cal F}$, is defined as the pullback of the current to the surface $r=0$.
As usual, this is defined by evaluating the pullback at some cutoff surface $r = r_\epsilon$  and taking the limit $r_\epsilon \to 0$, so we write
\begin{equation} \label{flux}
{\cal F} = \lim_{r_\epsilon \to 0} \frac{i}{2} \int_{r = r_\epsilon} d^{d_s} x d \xi \sqrt{\gamma} n^\mu j_\mu ,
\end{equation}
where $n_\mu$ the unit outward-pointing normal to the boundary.  The overall factor of $i/2$ is purely conventional.

As mentioned above, for constant perturbations the flux turns out to be finite. In the vector sector we find
\begin{align}
\nonumber
 	 {\cal F} =  - i \int_{r = 0} d^{d_s} x d \xi  & \bigg [  H_{\xi i}^{(-)} \wedge (2 H_{t i}^{(+)} - (z-1) s_i^{(+)}) + 2 H_{t i}^{(-)} \wedge H_{\xi i}^{(+)}    \\ 
 	 & + s_i^{(-)} \wedge (2 H_{t i}^{(+)} +  \frac{(z-1)(z+2)}{z} s_i^{(+)}) \bigg] ,
\end{align}
where  $A\wedge B=A_1 B_2-A_2 B_1$, where $1, 2$ label the two linearised solutions which define the current. This enables us to identify, up to an overall normalization which we neglect for simplicity,
\begin{equation}
\langle P_i \rangle = 2 H_{\xi i}^{(+)}, \quad  \langle T_i^\xi \rangle = 2 H_{ti}^{(+)} - (z-1) s_i^{(+)}, \quad \langle E^i \rangle = 2 H_{ti}^{(+)} + \frac{(z-1)(z+2)}{z} s_i^{(+)}.
\end{equation}

For non-zero $k_\xi$, $\omega$, the solutions of the linearised equations of motion can be given in a power series expansion; since the equations involve only $k_\xi \omega$ and $k_\xi^2$, this will be of the same form as in \eqref{tensor exp}. The interesting new feature here is that because of the different structure of the equations (we now need to solve \eqref{Hti xi kxi omega}), there is a reduction in the number of independent mode solutions. Solving \eqref{Hti xi kxi omega} at leading order implies a relation among the coefficients,
\begin{equation}\label{rel vectors 1}
	 k_\xi [ 2 H_{t i}^{(+)} - (z-1) s_i^{(+)} ] +2 \omega H^{(+)}_{\xi i} =0,
\end{equation}
which corresponds to the Ward identity
\begin{equation}
	\partial_t P_i + \partial_\xi T_i^\xi = 0,
\end{equation}
confirming our identification of the linearised solutions with the vevs.

\subsubsection{Scalar modes}

We consider now the scalar modes $H_{tt}$, $H_{t \xi}$, $H_{\xi \xi}$, $k$, $s_r$, $s_t$, $s_\xi$. They are governed by the equations
\begin{align}
\nonumber
0 =&\, \left(\frac{ k_\xi z}{2} + \omega z r^{2 z-2}\right) H_{\xi \xi}   - k_\xi z k +  (z-2) \left( k_\xi  + \omega r^{2 z-2} \right) s_\xi
+ k_\xi r s_t' -  k_\xi z s_t
\\\label{eq sc 1 kxi w}
&\, +( k_\xi  + \omega  r^{2 z-2} ) r s_\xi' - i \left( k_\xi^2 r^{2-z} + 2  k_\xi \omega  r^z + z (z+2) r^{z-2}\right) s_r ,
\\\nonumber
	0=&\, -\frac{1}{2} r (z+2) H_{\xi \xi}' + \left(\frac{1}{2} \omega^2 r^{2 z} - z^2 + z \right) H_{\xi \xi}
	- 3 r H_{t\xi}' - r^2 k_\xi \omega H_{t \xi}  + \frac{1}{2} k_\xi ^2 H_{tt} r^{4-2 z}
\\\label{eq sc 4 kxi w}
	&\, -3 r k' -  \left(k_\xi ^2 r^{4-2 z} + 2 k_\xi \omega r^2  \right) k
	+ (z-1) [ r s_\xi' + 2  z s_\xi -i k_\xi r^{2-z}  s_r ],
\\\nonumber
0=&\,	\frac{\omega}{2} r H_{\xi \xi}' + \left(\frac{1}{2} k_\xi (z-1) r^{2-2 z}+\frac{3}{2} \omega  (z-1)\right) H_{\xi \xi}  +
	\left(\frac{\omega }{2}-\frac{1}{2} k_\xi r^{2-2 z} \right) r H_{t \xi}'  \\\nonumber
&\,	- \frac{1}{2} k_\xi r^{2-2 z} r H_{tt}'  + \omega r k'  + (z-1) r^{2-2 z} k_\xi  [  H_{tt}  -  k +  s_t]
\\\label{eq sc 5 kxi w}
&\,	 - \omega (z-1) s_\xi - i \left(z^2+z-2\right) r^{-z} s_r,
\\
\label{eq sc 6 kxi w}
0=&\,	-\frac{1}{2} \omega r^{2 z-2} r H_{\xi \xi} ' - (z-1) \left(\omega   r^{2 z-2} + \frac{k_\xi}{2}\right) H_{\xi \xi}
	+\frac{1}{2} k_\xi r H_{t \xi} ' + k_\xi r k' ,
\\
\nonumber
0=&\,	\frac{1}{2} r^2 H_{\xi \xi}''  + r^2 H_{t \xi} '' + \frac{1}{2} r^2 H_{tt}''  + r^2 k''  + r \left(z-\frac{5}{2}\right) H_{\xi \xi}'  + 2 \left(z^2-3 z+2\right) H_{\xi \xi}
\\
\nonumber
&\,- (z+2) r H_{t \xi} ' + \frac{1}{2} (1-4 z) r H_{tt}' + 2 \left(z^2-1\right) H_{tt} + (z-4) r k'
\\
\nonumber
&\, - \left( k_\xi^2 r^{4-2 z} + 2 k_\xi \omega r^2  + \omega^2 r^{2 z} \right) k - 2 (z-1) r s_t' + 4 \left(z^2-1\right) s_t
\\\label{eq sc 7 kxi w}
&\,+  (1 - z) r s_\xi' + 4 (z-1) s_\xi + i  (z-1) \left( k_\xi r^{2-z} + 2 \omega r^z \right) s_r ,
\\\label{eq sc 9 kxi w}
0=&\,	r^2 H_{\xi \xi}'' + (4 z-7) r H_{\xi \xi}' + 4 \left(z^2-4 z+3\right) H_{\xi \xi} - 2 k_\xi^2 r^{4-2 z} k  ,
\\\nonumber
0=&\,
	r^2 H_{\xi \xi} '' + 2 r^2 H_{t \xi}'' + r^2 k''  + \left(\omega ^2 r^{2 z}-4 z+4\right) H_{\xi \xi}  - 6 r H_{t \xi}'
	+ k_\xi^2 r^{4-2 z} H_{tt}
\\\nonumber
&\, +  (z-4) r H_{\xi \xi}'  - 3 r k' - \left(k_\xi^2 r^{4-2 z} + 2 k_\xi \omega r^2  \right) k
\\\label{eq sc 10 kxi w}
&\,	- 2 k_\xi \omega r^2  H_{t \xi}  + 2 (z-1) [  i k_\xi r^{2-z} s_r  -  r s_\xi' + 4 s_\xi] .
\end{align}

In addition, we have equations
\begin{align}
\nonumber
0=&\,	r^2 s_\xi ''  +  2 (z-1) z H_{\xi \xi}  + k_\xi^2 r^{4-2 z} s_t +  (2 z-5) r s_\xi ' + z r H_{\xi \xi} '
\\\label{eq sc 2 kxi w}
&\,	- \left(k_\xi \omega r^2  + 8 z - 8 \right) s_\xi  + i  k_\xi r^{2-z}  (2 s_r  -  r s_r'),
\\\nonumber
0=&\,	r^2 s_\xi ''  +  r^2 s_t ''  - k_\xi \omega r^2 s_t
	+  \left(\omega ^2 r^{2 z}-2 z^2-2 z+4\right)  s_\xi   -   3 r  s_\xi'  - (2 z+1) r s_t'
\\\label{eq sc 3 kxi w}
&\,	+	\frac{1}{2} z r H_{\xi \xi} '  -  z r k'  +  2 i \left( k_\xi z r^{2-z} +  \omega  r^z  \right) s_r
	- i  \left( k_\xi r^{2-z} + \omega  r^{z} \right) r s_r' ,
\\\nonumber
0=&\,	\frac{1}{2} r^2 H_{\xi \xi} '' + \frac{1}{2} r^2 H_{t \xi} '' + r^2 k''  + \frac{3}{2} (z-2)  r H_{\xi \xi} '  + 2 \left(z^2-3 z+2\right) H_{\xi \xi}  +
\\\label{eq sc 8 kxi w}
&\,	-\frac{3}{2} r H_{t \xi} '   - 3 r k'  - \left( k_\xi^2 r^{4-2 z} + k_\xi \omega r^2 \right) k  .
\end{align}

For constant modes, (\ref{eq sc 1 kxi w},\ref{eq sc 5 kxi w},\ref{eq sc 6 kxi w}) are automatically satisfied if $s_r=0$, and (\ref{eq sc 2 kxi w},\ref{eq sc 3 kxi w},\ref{eq sc 8 kxi w}) are non-trivial equations. For general $k_\xi, \omega$, we solve (\ref{eq sc 1 kxi w}-\ref{eq sc 10 kxi w}), which imply (\ref{eq sc 2 kxi w},\ref{eq sc 3 kxi w},\ref{eq sc 8 kxi w}).

The solution for constant modes is
\begin{align}
	s_r =&\, 0, \\
\nonumber
	H_{tt} =& \, \frac{(3 z-2) s_\xi^{(-)}}{6 z}  r^{2-2 z} -2  s_t^{(0)}
	+ 2 r^{2z - 2} H_{tt}^{(-)} + r^{2z + 2} H_{tt}^{(+)}  \\
	&\,  + \frac{(6-5 z) H_{\xi \xi}^{(+)}}{4 (z-3) (z-2)} r^{6-2 z} + \frac{(6 k^{(4)} (z-4) + 5 s_\xi^{(4)} (z-1)(z+2))}{6 (z-3)} r^4 ,\\
		H_{t \xi} =&\,- ( H_{\xi \xi}^{(-)} + \frac{2}{3}  s_\xi^{(-)} ) r^{2-2 z} + s_t^{(0)} + H_{t \xi}^{(0)}
	- \frac{1}{2} H_{\xi \xi}^{(+)} r^{6 - 2z} +  \left( \frac{(z-1)(z+2)}{6} s_\xi^{(4)}  - k^{(4)} \right) r^4 ,\\
		H_{\xi \xi} =&\, 2 H_{\xi \xi}^{(-)} r^{2-2z} +  H_{\xi \xi}^{(+)} r^{6-2z} ,\\
		k  =&\, \frac{1}{3} s_\xi^{(-)}  r^{2- 2z} + 2 k^{(0)} + \frac{1}{2(3-z)} H_{\xi \xi}^{(+)} r^{6-2z}+ k^{(4)} r^4 ,\\
		s_\xi =&\, (H_{\xi \xi}^{(-)} +s_\xi^{(-)}) r^{2-2z} + s_\xi^{(4)} r^{4} +  \frac{z}{2(z-1)} H_{\xi \xi}^{(+)} r^{6-2z} ,\\
		s_t =&\, - \frac{1}{3}  s_\xi^{(-)}   r^{2 - 2z} + s_t^{(0)}  + s_t^{(+)} r^{2z+2}  + \frac{3 z  H_{\xi \xi}^{(+)} r^{6-2 z} }{4 (z-1)(z-3)} - \left( \frac{z  k^{(4)}}{2 (z-1)} +  \frac{(z+2)  s_\xi^{(4)} }{4} \right) r^{4} .
\end{align}
We have once again chosen the definition and normalization of the modes so that the $(0)$ and $(-)$ modes correspond to constant leading terms in the frame fields. The $r$-independent modes with a $(0)$ superscript correspond to sources for the diagonal components of the stress energy complex: $s_t^{(0)}$ is the constant part of $\delta \hat e^+_t$, so it is the source for the energy density $E$, $H_{t\xi}^{(0)}$ is the constant part of $\delta \hat e^-_\xi$, so it is the source for $P^\xi_\xi$, and $k^{(0)}$ is the constant part of $\delta \hat e^I_i$, so it is the source for the trace of the spatial stress tensor $T_i^i$. There is a single mode $H_{tt}^{(-)}$ of dimension $2z-2$, which comes from the constant part of $\delta \hat e^-_t$, so it is the source for the particle number density $P_\xi$. There are two modes of dimension $2-2z$, corresponding to sources for operators of dimension $2z+2$. The first is $H_{\xi\xi}^{(-)}$, which comes from the constant part of $\delta \hat e^+_\xi$, and hence corresponds to the source for the energy flux $E^\xi$. The second must be the source part of $\psi$, so we learn that this is dual to an operator $O$ of dimension $2z+2$. The source for this should not change   $\delta \hat e^+_\xi$, so we can identify this source as $s_\xi^{(-)}$. Note that as in the Lifshitz case, the source mode for this matter operator also appears in other fields, unlike the source modes for the stress tensor, whose appearance is constrained by the boundary diffeomorphism invariance.

We would again like to identify the remaining modes with the vevs of these operators. Dimensions alone suffices to fix $\langle P_\xi \rangle \sim H_{\xi\xi}^{(+)}$, to relate $\langle E \rangle$, $\langle T_i^i \rangle$ and $\langle P_\xi^\xi \rangle$ to $k^{(4)}$ and $s_{\xi}^{(4)}$, and to relate $\langle E^\xi \rangle$ and $\langle O \rangle$ to $H_{tt}^{(+)}$ and $s_t^{(+)}$.
To determine the relation we use the symplectic flux, which is calculated as in the vector sector. The symplectic flux is again finite and is given by
\begin{align}
\nonumber
	{\cal F} = i \int_{r = 0} d^{d_s} x d \xi & \bigg [ 
	 s_t^{(0)} \wedge \left (2 k^{(4)} + \frac{1}{3z} (z-1)(z^2-4z-6) s_\xi^{(4)} \right ) \\ 
\nonumber
	&\, +  H_{t\xi}^{(0)} \wedge \left (2 k^{(4)} + \frac{1}{3} (z-1)(z+2) s_\xi^{(4)}  \right) \\  
\nonumber
	&\,+ k^{(0)} \wedge \left ( - 4 k^{(4)} + \frac{2}{3} (z-1)(2z+1) s_\xi^{(4)} \right) - 2  H_{tt}^{(-)} \wedge H_{\xi\xi}^{(+)} \\ 
	&\,-  H_{\xi\xi}^{(-)} \wedge \left( 2 H_{tt}^{(+)} + \frac{2(z-1)}{z} s_t^{(+)} \right) - \frac{2 (z^2-1)}{z} s_\xi^{(-)} \wedge s_t^{(+)} \bigg ] .
\end{align}

This implies the identifications
\begin{equation}
\langle P_\xi \rangle = 2 H_{\xi\xi}^{(+)}, \quad \langle P_\xi^\xi \rangle = -2 k^{(4)} - \frac{1}{3} (z-1)(z+2) s_\xi^{(4)} ,
\end{equation}
\begin{equation}
\langle E \rangle = - 2 k^{(4)} - \frac{1}{3z} (z-1)(z^2-4z-6) s_\xi^{(4)},
\end{equation}
and
\begin{equation}
\langle T_i^i \rangle  = \langle T_1^1 \rangle + \langle T_2^2 \rangle =  4 k^{(4)} - \frac{2}{3} (z-1)(2z+1) s_\xi^{(4)}.
\end{equation}
The Ward identity from the scaling invariance is $z E + T_i^i + (2-z) P_\xi^\xi =0$, which is indeed satisfied by these vevs. The other vevs are
\begin{equation}
\langle E^\xi \rangle = 2 H_{tt}^{(+)} + \frac{2(z-1)}{z} s_t^{(+)}
\end{equation}
and
\begin{equation}
\langle O \rangle =  \frac{2(z^2-1)}{z} s_t^{(+)}.
\end{equation}

For a solution with non-zero $k_\xi$, $\omega$, there is a power series expansion of the same form as in \eqref{tensor exp}, with an extra factor of $k_\xi$ in $s_r$. As in the vector case, there is a reduction in the number of independent mode solutions because of the different structure of the equations of motion. There is a linear combination of (\ref{eq sc 1 kxi w},\ref{eq sc 5 kxi w}) which is independent of $s_r$. That and \eqref{eq sc 6 kxi w} imply the relations
\begin{align}
 	 k_\xi \left( k^{(4)}  + \frac{1}{6} (z-1)(z+2) s_\xi^{(4)}  \right ) -\omega H_{\xi \xi}^{(+)} & = 0 ,
\\
	 	-2  k_\xi \left (  H_{tt}^{(+)} + \frac{(z-1)}{z} s_t^{(+)} \right ) + \omega \left ( 2 k^{(4)} +
 	 	\frac{1}{3z} (z-1)(z^2 - 4z - 6) s_\xi^{(4)} \right ) &= 0;	
\end{align}
which correspond to the Ward identities
\begin{align}
	\partial_t P_\xi + \partial_\xi P^\xi_\xi &= 0 ,\\	\partial_t E + \partial_\xi E^\xi &= 0 .
\end{align}
This confirms our identification of the vevs in terms of the linearised modes. The identification of sources and vevs for the different operators is summarized in table 1. 

\begin{table}
\begin{center}
\begin{tabular}{|l||l|l|}
\hline
Operator & Source & Expectation value \\
\hline
$E$ & $s_t^{(0)}$ & $- 2 k^{(4)} - \frac{1}{3z} (z-1)(z^2-4z-6) s_\xi^{(4)}$ \\
$E^i$ & $s_i^{(-)}$ & $2 H_{ti}^{(+)} + \frac{(z+1)(z+2)}{z} s_i^{(+)}$ \\
$E^\xi$ & $H_{\xi\xi}^{(-)}$ & $2 H_{tt}^{(+)} + \frac{2(z-1)}{z} s_t^{(+)}$ \\
$P_i = P^i_\xi$ & $H_{ti}^{(-)}$ & $2 H_{\xi i}^{(+)}$ \\
$T^1_1 + T^2_2$ & $k^{(0)}$ & $4 k^{(4)} - \frac{2}{3} (z-1)(2z+1) s_\xi^{(4)}$ \\
$T^1_1 - T^2_2$, $T^1_2$ & $\bar H_{ij}^{(0)}$ & $\bar H_{ij}^{(4)}$ \\
$T^\xi_i$ & $H_{\xi i}^{(-)}$ & $2 H_{ti}^{(+)} - (z-1) s_i^{(+)}$ \\ 
$P_\xi$ & $H_{tt}^{(-)}$ & $2 H_{\xi\xi}^{(+)}$ \\ 
$P^\xi_\xi$ & $H_{t\xi}^{(0)}$ & $-2 k^{(4)} - \frac{1}{3} (z-1)(z+2) s_\xi^{(4)}$ \\ 
$O$ & $s_\xi^{(-)}$ & $\frac{2(z^2-1)}{z} s_t^{(+)}$ \\
\hline
\end{tabular}
\end{center}
\caption{The identification of linearised modes with sources and vevs for the operators in the dual field theory.}
\end{table}

\subsubsection{Linearised solutions with spatial dependence}
\label{spatial}

The most general linearised solutions include spatial dependence. Considering a single Fourier mode in all boundary directions, we can use the rotation symmetry to orient the spatial coordinates so that the spatial momentum is along the $x$ direction, so the coordinate dependence in all modes is $e^{i \omega t + i k_\xi \xi + i k_x x}$. Then the modes split up into the scalar modes $H_{tt}$, $H_{t\xi}$, $H_{\xi\xi}$, $H_{tx}$, $H_{\xi x}$, $H_{xx}$, $H_{yy}$ $s_t$, $s_x$, $s_r$ and the vector modes $H_{ty}$, $H_{\xi y}$, $H_{xy}$, $s_y$. As in the discussion with no spatial dependence, these all have an expansion in powers of $k_\xi \omega r^2$ and $k_x^2 r^2$. The leading terms take the same form as for the constant modes above.  

The equations of motion in the vector sector are
\begin{align}
\label{E45}
&rk_x[r^z \omega H_{\xi y}+k_\xi r^{2-z}(H_{ty}+H_{\xi y})]-(k_\xi^2 r^{4-2z}+2k_\xi\omega r^2)H_{xy}-3r H_{xy}'+r^2 H_{xy}''=0,
\\
\label{M5}
&z(z-1)H_{\xi y}-k_x^2 r^2 s_y - (k_\xi^2 r^{4-2z}+2k_\xi\omega r^2 + (z-1)(z+3))s_y+r(zH_{\xi y}'-3s_y'+rs_y'')=0,
\\
\label{E15}
&k_\xi [(z-1)(H_{ty}-H_{\xi y}+2s_y)-r(H_{ty}+H_{\xi y})']-\omega r^{2z}[(z-1)H_{\xi y}+rH_{\xi y}']-k_x r^z H_{xy}'=0,
\\\nonumber
&k_x (k_\xi r^{3-z}H_{xy}-k_x r^2H_{\xi y})+k_\xi^2 r^{4-2z}H_{ty}+((z-1)(z-5)-k_\xi\omega r^2)H_{\xi y}
\\\label{E35}
& \hspace{9.5cm}+ r((2z-5)H_{\xi y}' + rH_{\xi y}'')=0,
\\\nonumber
&k_xr(r^{z}\omega H_{xy}-k_x r H_{ty})+(k_\xi\omega r^2 - k_\xi^2 r^{4-2z} - (z-1)(z+3))H_{ty} +2(z-1)(z+3)s_y
\\\label{E25}
& \hspace{.3cm} + (k_\xi\omega r^2 + \omega^2 r^{2z}-2(z-1)^2)H_{\xi y}+r(rH_{ty}''-2(z-1)(H_{\xi y}+s_y)'-(1+2z)H_{ty}')=0.
\end{align}
We can solve these equations order by order in $k_\xi \omega$ and $k_x^2$. The subleading components determine the subleading terms in the expansion of the fields. But there are also additional constraints on the leading terms, corresponding to the expected Ward identities. Equation (\ref{E15}) gives at leading order
\begin{equation}
\label{WE15}
k_\xi [ 2H_{ty}^{(+)}-(z-1)s_y^{(+)}] + 2 \omega H_{\xi y}^{(+)} + k_x \bar{H}_{xy}^{(4)} =0
\end{equation}
which corresponds to the Ward identity
\begin{equation}
\partial_t P_y + \partial_\xi T^\xi_y + \partial_x T^x_y =0.
\end{equation}

In the scalar sector, the equations of motion are
\begin{align}
\nonumber
0=&\, 2k_x r(k_\xi r^{2-z} + \omega r^z)(H_{tx}+H_{\xi x})+k_x^2 r^2\Big(\frac{1}{2} H_{tt}-\frac{1}{2}H_{\xi\xi}-H_{t\xi}-k \Big)
-2(1-z^2)H_{tt}
\\\nonumber
&\, +2(2-3z+z^2)H_{\xi\xi}-(k_\xi r^{2-z}+\omega r^z)^2 k + i(z-1)(k_\xi r^{2-z} + 2\omega r^z)s_r -4(1-z^2)s_t
\\\nonumber
&\, -4(1-z)s_\xi-(z+2)rH_{t\xi}'+\frac{1}{2}(2z-5)rH_{\xi\xi}'+(z-4)rk' -(z-1)rs_\xi '
\\
&\, + r[(1-4z)H_{tt}' - 2(z-1)s_t '] + \frac{1}{2} r^2 H_{tt}'' + r^2H_{t\xi}'' + \frac{1}{2} r^2 H_{\xi\xi}'' + r^2 k'',
\label{E22}
\\
\nonumber
0=&\, k_x r \Big( k_\xi r^{2-z}H_{tx} + (2k_\xi r^{2-z}+\omega r^z)H_{\xi x} \Big) -k_x^2 r^2(H_{t\xi}+H_{\xi\xi} + k) + 4(z^2-3z+2)H_{\xi\xi}
\\
&\, -2(k_\xi^2 r^{4-2z} + \omega k_\xi r^2)k -3r \Big( H_{t\xi}'-(z-2)H_{\xi\xi}' + 2k' \Big) + r^2 H_{t\xi}''+ r^2 H_{\xi\xi}'' + 2r^2 k'' ,
\label{E23}
\\
\nonumber
0=&\, rk_x[\omega r^z(H_{\xi\xi} + k)-2i (z-1) s_r-k_\xi H_{tt}+(\omega r^z-k_\xi r^{2-z})H_{t\xi}]
\\\nonumber
&\, +(k_\xi^2 r^{4-2z} + k_\xi \omega r^2 - (z-1)(z+3))H_{tx}-(k_\xi\omega r^2+\omega^2 r^{2z} - 2(z-1)^2)H_{\xi x}
\\
&\, -2(z-1)(z+3)s_x + (2z+1)rH_{tx}'+2r(z-1)(H_{\xi x}' + s_x')-r^2 H_{tx}'',
\label{E24}
\\
\nonumber
0=&\,-2 k_x k_\xi r^{3-z} H_{\xi x}+k_x^2 r^2 H_{\xi\xi} - 4(z-1)(z-3)H_{\xi\xi}+2k_\xi^2 r^{4-2z} k+(7-4z)rH_{\xi\xi}'
\\
&\, -r^2 H_{\xi\xi}'',
\label{E33}
\\
\nonumber
0=&\,k_x r \Big( k_\xi r^{2-z}(H_{t\xi}+k) - \omega r^z H_{\xi\xi} \Big)-k_\xi^2 r^{4-2z} H_{tx} +  \Big( k_\xi \omega r^2 - (z-1)(z-5) \Big) H_{\xi x}
\\
&\, +(5-2z)rH_{\xi x}' - r^2 H_{\xi x}'',
\label{E34}
\\
\nonumber
0=&\, r^2 k'' + r^2 H_{\xi\xi}'' + 2 r^2 H_{t\xi}''-(z-1)rs_\xi '-3rk' + (z-4)r H_{\xi\xi}'-6r H_{t\xi}'+8(z-1)s_\xi
\\\nonumber
&\, + 2i (z-1) k_\xi r^{2-z} s_r - (k_\xi^2 r^{4-2z} + 2\omega k_\xi r^2)k + \Big( 4(1-z)+ \omega^2 r^{2z} \Big) H_{\xi\xi} - 2k_\xi\omega r^2 H_{t\xi}
\\
&\, +k_\xi^2 r^{4-2z} H_{tt},
\label{E44}
\\
\nonumber
0=&\, 2k_x r \Big( k_\xi r^{2-z}(H_{tx}+2 H_{\xi x}) + \omega r^z H_{\xi x} \Big) -r^2 k_x^2 (2H_{tx}+H_{\xi\xi})+r^2 k'' + r^2 H_{\xi\xi}''
\\\nonumber
&\, -2(z-1)r s_\xi ' - 3rk' + (z-4) r H_{\xi\xi}' - 6r H_{t\xi}'+8(z-1)s_\xi + 2i(z-1)k_\xi r^{2-z} s_r
\\
&\, - (k_\xi^2 r^{4-2z} + 2\omega k_\xi r^2)k + \Big( 4(1-z) + \omega^2 r^{2z} \Big) H_{\xi\xi} - 2 k_\xi \omega r^2 H_{t\xi} + k_\xi^2 r^{4-2z} H_{tt},
\label{E55}
\\
\nonumber
0=&\, k_x k_\xi r^{3-z} s_x - k_x^2 r^2 s_\xi + 2z(z-1)H_{\xi\xi} + 2i k_\xi r^{2-z} s_r + k_\xi^2 r^{4-2z} s_t - \Big( 8(z-1)+k_\xi \omega r^2 \Big) s_\xi
\\
&\, rz H_{\xi\xi}' - i k_\xi r^{3-z} s_r ' + (2z-5)r s_\xi ' + r^2 s_\xi '',
\label{M2}
\\
\nonumber
0=&\,s k_x r(k_\xi r^{2-z} + \omega r^z) s_x - 2 k_x^2 r^2 (s_\xi + s_t)+ 4i (z k_\xi r^{2-z} + \omega r^z)s_r
\\\nonumber
&\, + \Big( 2\omega^2 r^{2z} - 4(z-1)(z+2)\Big) s_\xi + zr H_{\xi\xi}' - 2z r k' - 2i (k_\xi r^{2-z} + \omega r^z) r s_r ' - 6r s_\xi ' - 2 k_\xi \omega r^2 s_t
\\
&\, - 2(1+2z) r s_t' + 2r^2 s_t '' + 2r^2 s_\xi '',
\label{M3}
\\
\nonumber
0=&\,k_x r[2i s_r + k_\xi r^{2-z} s_t + (k_\xi r^{2-z} + r^z \omega)s_\xi - i r s_r '] + z(z-1)H_{\xi x}
\\
&-(k_\xi^2 r^{4-2z} + 2k_\xi \omega r^2 + (z-1)(z+3))s_x + z r H_{\xi x}' - 3r s_x ' + r^2 s_x '',
\label{M4}
\end{align}
and additionally,
\begin{align}
\nonumber
0=&\, k_\xi^2 r^{4-2z}H_{tt}+2k_\xi k_x r^{3-z} H_{tx}-2 (k_x^2 + k_\xi \omega)r^2 H_{t\xi}+2rk_x(k_\xi r^{2-z}+\omega r^z)H_{\xi x}
\\
\nonumber
&\,-\Big( k_x^2 r^2 +2z(z-1) + \omega^2 r^2\Big) H_{\xi\xi}-\Big(2k_\xi^2 r^{4-2z} + r^2(k_x^2 + 4k_\xi \omega) \Big) k
\\
&\,-2ik_\xi(z-1)r^{2-z}s_r + 4z(z-1)s_\xi -6rH_{t\xi}'-(2+z)rH_{\xi\xi}'-6rk'+2(z-1)rs_\xi ',
\label{E11}
\\
\nonumber
0=&\, k_x r\Big((z-1)(H_{tx}+H_{\xi x})-rH_{tx}'\Big)+(z-1)(k_\xi r^{2-z} + 3\omega r^z)H_{\xi\xi}-2k_\xi(z-1)r^{2-z}k
\\
\nonumber
&\,-2i(z-1)(z+2)s_r-2(z-1)\omega r^z s_\xi+k_\xi r^{2-z}\Big( 2(z-1)(H_{tt}+s_t)-rH_{tt}' \Big)
\\
&\,+(-k_\xi r^{2-z} + \omega r^z)rH_{t\xi}'+\omega r^{z+1}(H_{\xi\xi}'+k'),
\label{E12}
\\
\nonumber
0=&\,-k_x r \Big( (z-1)H_{\xi x}+rH_{\xi x}' \Big)-(z-1)(k_\xi r^{2-z} + 2\omega r^z)H_{\xi\xi}+k_\xi r^{3-z} H_{t\xi}'-r\omega r^z H_{\xi\xi}'
\\
&\,+2k_\xi r^{3-z}k',
\label{E13}
\\
\nonumber
0=&\, k_\xi (z-1) r^{2-z}(H_{tx}-H_{\xi x}+2s_x)-(z-1)\omega r^z H_{\xi x}-k_\xi r^{3-z} H_{tx}'-(k_\xi r^{2-z} + \omega r^z)rH_{\xi x}'
\\
&\, +k_x r\Big((z-1)(H_{\xi\xi}+2s_\xi)+2r(H_{t\xi}'+H_{\xi\xi}'+k' ) \Big),
\label{E14}
\\
\nonumber
0=&\,k_x r(2z H_{\xi x}+ 2 s_x + r s_x ') - 2i k_x^2 r^2 s_r + (k_\xi r^{2-z} + 2\omega r^z)zH_{\xi\xi}-2z k_\xi r^{2-z} k
\\\nonumber
&\, -2i \Big( k_\xi^2 r^{4-2z} + z(z+2) + 2k_\xi \omega r^2 \Big) s_r + 2(z-2)(k_\xi r^{2-z} + \omega r^z) s_\xi + 2k_\xi r^{2-z}(-zs_t + r s_t ')
\\
&\, + 2(k_\xi r^{2-z} + \omega r^z) s_\xi '.
\label{M1}
\end{align}
Again, the constraints corresponding to the Ward identities are modified. Equation (\ref{E14}) gives\begin{equation}
\label{WE14}
k_\xi [ 2H_{tx}^{(+)}-(z-1)s_x^{(+)}] + 2 \omega H_{\xi x}^{(+)} + k_x [ 2k^{(4)} - \frac{1}{3}(z-1)(2z+1)] =0,
\end{equation}
 which corresponds to the Ward identity
\begin{equation}
\partial_t P_x + \partial_\xi T^\xi_x + \partial_x T^x_x =0;
\end{equation}
(\ref{E13}) gives
\begin{equation}
k_\xi \Big( k^{(4)} + \frac{1}{6}(z-1)(z+2)s_\xi^{(4)}\Big) - \omega H_{\xi\xi}^{(+)} - k_x H_{\xi x}^{(+)} =0,
\end{equation}
which corresponds to the Ward identity
\begin{equation}
\partial_t P_\xi + \partial_\xi P^\xi_\xi + \partial_x P^x_\xi =0;
\end{equation}
and there is a  linear combination of (\ref{M1}) and (\ref{E12}) which is independent of $s_r$ which gives
\begin{equation}
k_\xi \Big( 2H_{tt}^{(+)} + \frac{2(z-1)}{z} s_t^{(+)} \Big) - \omega \Big( 2k^{(4)} + \frac{1}{3z} (z-1)(z^2 - 4z - 6) \Big)
+ k_x \Big( 2H_{tx}^{(+)} + \frac{1}{z}(z-1)(z+2)s_x^{(+)} \Big) = 0
\end{equation}
which corresponds to the Ward identity
\begin{equation}
\partial_t E + \partial_\xi E^\xi + \partial_x E^x =0.
\end{equation}
Thus the full linearised perturbations behave as we expect.

\subsection{Linearised solutions for $d_s=0$}

The solution for other values of $d_s$ is qualitatively similar to the one discussed above, but the three-dimensional bulk is a special case. In this case there are no spatial dimensions. Hence the previous analysis of the spatially independent modes corresponds to the general analysis in this case, and there are no vector or tensor modes, so the structure is similar to the scalar mode analysis in $d_s=2$. There will be no field $k$ in this case, corresponding to the absence of the trace of the spatial stress tensor.

The solution for the constant modes is
\begin{align} \label{3dc}
	s_r &= 0 ,\\
	H_{tt} &=  -2 s_t^{(0)} + H_{tt}^{(+)} r^{2 z} + 2 H_{tt}^{(-)} r^{2 z-2} - \frac{z H_{\xi \xi}^{(+)} r^{4-2 z}} {2(z-2)(2z-3)}
	+\frac{2 z (z-1) r^2 s_\xi^{(2)}}{(z-2)}  ,\\
		H_{t \xi} &=  s_t^{(0)}+ H_{t\xi}^{(0)} - H_{\xi \xi}^{(-)}  r^{2-2 z} - \frac{(z-1) H_{\xi \xi}^{(+)}  r^{4-2 z}}{2 (z-2)}+ z (z-1) r^2 s_\xi^{(2)} ,\\
		H_{\xi \xi} &=  2H_{\xi \xi}^{(-)} r^{2-2 z}  +  H_{\xi \xi}^{(+)} r^{4-2 z} ,\\
		s_\xi &= (H_{\xi \xi}^{(-)}+ s_\xi^{(-)}) r^{2-2 z} + r^2 s_\xi^{(2)} + \frac{z H_{\xi \xi}^{(+)} r^{4-2 z}}{2 (z-1)} ,\\
		s_t &=  s_t^{(+)} r^{2 z} + s_t^{(0)}  -\frac{z s_\xi^{(-)}  r^{2-2 z}}{4 z-2} + \frac{ z H_{\xi \xi}^{(+)}  r^{4-2 z}}{4 (z-2)(z-1)}
		 -  \frac{z}{2} r^2 s_\xi^{(2)}  .
\end{align}
As in the previous case, the $r$-independent modes correspond to sources for the stress energy complex: $s_t^{(0)}$ is the constant part of $\delta \hat e^+_t$, so it is the source for the energy density $E$, and $H_{t\xi}^{(0)}$ is the constant part of $\delta \hat e^-_\xi$, so it is the source for $P^\xi_\xi$. There is a single mode $H_{tt}^{(-)}$ of dimension $2z-2$, which comes from the constant part of $\delta \hat e^-_t$, so it is the source for the particle number density $P_\xi$. The two modes of dimension $2-2z$ are $H_{\xi\xi}^{(-)}$, which comes from the constant part of $\delta \hat e^+_\xi$, and hence corresponds to the source for the energy flux $E^\xi$, and $s_\xi^{(-)}$, which is the source for an operator $O$ of dimension $2z+2$.

We would again like to identify the remaining modes with the vevs of these operators. Dimensions alone suffice to fix $\langle P_\xi \rangle \sim H_{\xi\xi}^{(+)}$, to relate $\langle E \rangle$ and $\langle P_\xi^\xi \rangle$ to $s_{\xi}^{(2)}$, and to relate $\langle E^\xi \rangle$ and $\langle O \rangle$ to $H_{tt}^{(+)}$ and $s_t^{(+)}$. The symplectic flux is

\begin{align}
\nonumber
     {\cal F} =&\, -i \int_{r = 0} d^{d_s} x d \xi \bigg[  H_{\xi \xi}^{(-)} \wedge  H_{tt}^{(+)} + H_{tt}^{(-)}\wedge H_{\xi \xi}^{(+)}  \\
	&\,+ 2 (z-1) s_\xi^{(-)} \wedge  s_t^{(+)}  - z (z-1)H_{t \xi}^{(0)}\wedge s_\xi^{(2)} - (z-1)(z-2) s_t^{(0)} \wedge s_\xi^{(2)} \bigg] .
\end{align}
This enables us to identify the vevs
\begin{equation}
\langle P_\xi \rangle =  H_{\xi\xi}^{(+)}, \quad \langle P_\xi^\xi \rangle = -z(z-1) s_\xi^{(2)} ,
\end{equation}
\begin{equation}
\langle E \rangle = -(z-1)(z-2) s_\xi^{(2)},
\end{equation}
which indeed satisfy the Ward identity from the scaling invariance, which is $z E + (2-z) P_\xi^\xi =0$,
\begin{equation}
\langle E^\xi \rangle = H_{tt}^{(+)},
\end{equation}
and
\begin{equation}
\langle O \rangle =  2(z-1) s_t^{(+)}.
\end{equation}

For non-zero $k_\xi$, $\omega$, there will be Ward identities $\partial_t E + \partial_\xi E^\xi= 0$, $\partial_t P_\xi + \partial_\xi P_\xi^\xi =0$ and $zE + (2-z) P_\xi^\xi =0$, which leave us with just one free vev in the stress energy complex.

\subsubsection{Comparison to previous work}

In \cite{Costa:2010cn}, a full linearised analysis for $z <2$ and $d_s=0$ was carried out. They write the Schr\"odinger metric in a different radial coordinate, $\rho = r^2$, and introduce $\sigma$ as discussed in the introduction by rescaling the boundary coordinates, $u^2 = -t^2/\sigma^2$, $v^2 = -\sigma^2 \xi^2$, so the Schrodinger metric becomes
\begin{equation}
ds^2 = \frac{d\rho^2}{4 \rho^2} + \frac{2 du \,dv}{\rho} + \frac{\sigma^2 du^2}{\rho^z}.
\end{equation}
Their focus is on $z <1$, where $\sigma^2 >0$; for the range $z >1$ we are interested in we need $\sigma^2 < 0$. We will henceforth set $\sigma^2 = -1$; then their $b$ is identical to our $\alpha$. The linearised perturbations are written as
\begin{equation}
A^{(1)}_\mu = a_\mu, \quad g^{(1)}_{ab} = \rho^{-1} h_{ab},
\end{equation}
where $\mu = u,v,r$, $a,b = u,v$. Relative to our definitions above,
\begin{equation}
a_u = \alpha  \rho^{-z/2 } s_t, \quad a_v = \alpha \rho^{z/2-1} s_\xi, \quad a_r = \alpha \rho^{-1/2} s_r,
\end{equation}
and
\begin{equation}
h_{uu} = \rho^{1-z} H_{tt}, \quad h_{uv} = H_{t\xi}, \quad h_{vv} = \rho^{z-1} H_{\xi\xi}.
\end{equation}
In \cite{Costa:2010cn}, the perturbation is decomposed into a part which only affects the metric $h_{ab}$ and a $V$ mode which enters in both $a_\mu$ and $h_{ab}$ (following the decomposition into T and X modes in \cite{Guica:2010sw}).

To relate to our analysis above, we will consider the case where the modes are constant in the boundary directions, so $h_{ab}$, $a_\mu$ are functions only of $\rho$. As in our analysis, this implies that $a_\rho=0$. The metric can be written as
\begin{equation}
h_{vv} = h_{(0)vv} + \rho h_{(2)vv},  \quad h_{uv} = h_{(0)uv} + \rho h_{(2)uv} - \frac{1}{2} \rho^{1-z} h_{(0)vv} - \frac{(1-z)}{2(2-z)} \rho^{2-z} h_{(2)vv},
\end{equation}
\begin{equation}
h_{uu} = h_{(0)uu} + \rho h_{(2)uu}  - \frac{z}{4(1-2z)} \rho^{2-2z} h_{(0)vv}  - \frac{z}{4(3-2z)} \rho^{3-2z} h_{(2)vv}  - \frac{1}{2-z} \rho^{2-z} h_{(2)uv} +   h_{uu}^V,
\end{equation}
where
\begin{equation}
\partial_\rho^2 h_{uu}^V = \frac{z\alpha}{2} \rho^{1-z} \partial_\rho ( \rho^{-z/2} a_v) + z \alpha \partial_\rho ( \rho^{-z/2} a_u).
\end{equation}
In solving this equation, we will take $h_{uu}^V$ to have no constant or linear pieces in $\rho$, so that $h_{(0)uu}$ and $h_{(2)uu}$ represent the whole of the $\rho^0$ and $\rho$ coefficients. The vector field satisfies
\begin{equation}
\rho^{z/2} \partial_\rho [ \rho^{1-z} \partial_\rho ( \rho^{z/2} a_v)] = - \frac{z \alpha}{2} \rho^{-z/2}h_{(2)vv} ,
\end{equation}
\begin{equation}
\rho^{z/2} \partial_\rho [ \rho^{1-z} \partial_\rho ( \rho^{z/2} a_u)] = -(1-z) \rho^{1-z} \partial_\rho a_v - \frac{z(1-z) \alpha}{4} \rho^{-3z/2} h_{(0)vv} + \frac{z^2\alpha}{4}  \rho^{1-3z/2} h_{(2)vv};
\end{equation}
and there is a single constraint for constant solutions,
\begin{equation} \label{constc}
-4 h_{(2)uv} - 2z \rho^{1-z} h_{(2)vv} + 2 z \alpha \rho^{1-z} \partial_\rho ( \rho^{z/2} a_v) = 0.
\end{equation}
Solving this system of equations, we find that
\begin{align}
a_v = &\, \alpha_v \rho^{-z/2} + \beta_v \rho^{z/2} + \frac{z \alpha}{2(z-1)} h_{(2)vv} \rho^{1-z/2},
\\\nonumber
a_u  =&\, \alpha_u \rho^{-z/2} + \beta_u \rho^{z/2} - \frac{z}{4z-2} \alpha_v \rho^{1-3z/2}  - \frac{z}{2} \beta_v \rho^{1-z/2} + \frac{z\alpha}{4(2z-1)} h_{(0)vv} \rho^{1-3z/2}
\\ &\, + \frac{z \alpha}{4(z-2)(z-1)} h_{(2)vv} \rho^{2-3z/2},
\\\nonumber
h_{uu} =&\, h_{(0)uu} + \rho h_{(2)uu} - \frac{z}{2(z-2)(2z-3)} h_{(2)vv} \rho^{3-2z} - \frac{1}{(2-z)} h_{(2)uv} \rho^{2-z}
\\ &\, - \frac{z \alpha}{(z-1)} \alpha_u \rho^{1-z}  - \frac{z^2 \alpha}{2(2-z)} \beta_v \rho^{2-z},
\end{align}
where the constraint \eqref{constc} implies that $h_{(2)uv} = \frac{z^2 \alpha}{2} \beta_v$. The constants $\alpha_{u,v}$, $\beta_{u,v}$ correspond to the $V$ mode solutions of \cite{Costa:2010cn}.

Comparing to our constant solutions, we see that we can identify the sources
\begin{equation}
h_{(0)vv} = 2 H_{\xi\xi}^{(-)}, \quad h_{(0)uv} = s_t^{(0)} + H_{t\xi}^{(0)}, \quad h_{(0)uu} = 2 H_{tt}^{(-)},
\end{equation}
\begin{equation}
\alpha_v = \alpha (H_{\xi\xi}^{(-)} + s_\xi^{(-)}), \quad \alpha_u = \alpha s_t^{(0)};
\end{equation}
and vevs
\begin{equation}
h_{(2)vv} = H_{\xi\xi}^{(+)}, \quad h_{(2)uv} = z(z-1) s_\xi^{(2)}, \quad h_{(2)uu} =  H_{tt}^{(+)},
\end{equation}
\begin{equation}
 \quad \beta_v = \alpha s_\xi^{(2)}, \quad \beta_u = \alpha s_t^{(+)}.
\end{equation}
As we might have expected, while the sources for the momentum density and flux appear only in the metric modes, the sources for the energy density and flux appear also in the $V$ modes. The vev mode $s_\xi^{(2)}$ also appears in the $V$ modes. The source and vev for the operator $O$ appear only in the $V$ modes and not in the metric modes. The constraint \eqref{constc} imposes the trace Ward identity.

When we go beyond constant modes, there will be subleading terms in the $V$ modes, determined by solving the equations in \cite{Costa:2010cn}. There are also additional constraints; there is a constraint
\begin{equation}
\partial_v h_{(2)uv} = \partial_u h_{(2)uu},
\end{equation}
which corresponds precisely to the expected Ward identity $\partial_t P_\xi + \partial_\xi P_\xi^\xi = 0$, and a constraint
\begin{align}
\partial_v \partial_\rho h_{uu} =&\, \partial_u h_{(2) uv} + \frac{z}{2} \alpha \rho^{-z/2} ( -2 \partial_u a_v + 2z a_\rho - 4 \rho \partial_\rho a_\rho - \rho^{1-z} \partial_v a_v)
\\ &\, + \frac{z \alpha^2}{4} \rho^{-z/2} ( 4 \partial_u h_{(0)vv} + 4 \rho \partial_u h_{(2)vv} + \rho^{1-z} ( \partial_v h_{(0)vv} + \rho \partial_v h_{(2)vv} )).\nonumber
\end{align}
The $\rho$ derivative of this constraint vanishes by virtue of the other equations of motion; the constant part gives
\begin{equation}
\partial_v  h_{(2)uu} = \partial_u h_{(2) uv} - z \alpha \partial_u \beta_v,
\end{equation}
which is precisely the expected Ward identity $\partial_t E + \partial_\xi E^\xi = 0$.

Thus our solution is consistent with the one in \cite{Costa:2010cn}, but our frame perspective offers a different physical interpretation with a new organisation of the sources and vevs. We agree with \cite{Costa:2010cn} on the split of the linearised solutions into sources and vevs, but we give a different physical interpretation to these sources and vevs in terms of operators in the field theory.

\section{Asymptotic expansion for $z<2$}
\label{asymp}

In this section, we want to go beyond the linearised analysis by showing that solutions of the bulk equations of motion exist for arbitrary boundary data. To do so, we will solve the equations of motion in an asymptotic expansion: that is, we work at large $r$, and solve the equations in an expansion in powers of $r$. We will follow closely the treatment of the asymptotic expansion for asymptotically Lifshitz spacetimes in \cite{Ross:2011gu}, using a radial Hamiltonian framework to analyse the equations. In the course of demonstrating the existence of this asymptotic expansion, we will also see that when the asymptotic expansion exists we can cancel the divergent terms in the action in the usual way by adding appropriate local counterterms determined by the boundary data.

The action we consider is a massive vector theory, which is the same as the theory considered in \cite{Ross:2011gu}, so the equations are the same. However \cite{Ross:2011gu} considered a four-dimensional bulk, whereas our main interest here is a five-dimensional bulk, so some dimension-dependent factors are different. For generality, we write the equations for general $d_s$. By taking the trace, we can rewrite \eqref{eins eqs} as
\begin{equation}\label{Riccigenerald}
R_{\mu\nu} = \frac{2}{d-2}\Lambda g_{\mu\nu} + \frac{1}{2}F_{\mu\lambda}F_\nu^{\ \lambda}
- \frac{1}{4(d-2)}F_{\lambda \rho}F^{\lambda\rho}g_{\mu\nu} + \frac{1}{2}m^2 A_\mu A_\nu,
\end{equation}
where $d = d_s+3$ is the dimension of the bulk spacetime.
The Gauss-Codazzi equations on a surface of constant $r$ are then
\begin{align}\label{Simon26}
\dot K_{\alpha\beta} +K K_{\alpha\beta}-2K_{\alpha\gamma}K^\gamma_{\ \beta} = & R_{\alpha\beta}-\frac{2}{d-2}\Lambda h_{\alpha\beta}-\frac{1}{2}F_{\alpha\gamma}F_\beta^{\ \gamma}+\frac{2}{8(d-2)}h_{\alpha\beta}F_{\gamma\delta}F^{\gamma\delta}
\nonumber\\
&-\frac{1}{2}\pi_\alpha\pi_\beta
+\frac{2}{4(d-2)}h_{\alpha\beta}\pi_\gamma \pi^\gamma-\frac{1}{2}m^2 A_\alpha A_\beta,
\end{align}
\begin{equation}\label{Simon27}
\dot\pi^\alpha + K \pi^\alpha+ \nabla_\beta F^{\beta\alpha}=m^2 A^\alpha
\end{equation}
and the constraints become
\begin{equation}\label{Simon28}
 \nabla_\alpha K^\alpha_{\ \beta}-\nabla_\beta K^\alpha_{\ \alpha} = \frac{1}{2}F_{\beta \alpha}\pi^\alpha + \frac{1}{2}m^2 A_\beta  A_n,
\end{equation}
\begin{align}\label{Simon29}
&K^2-K_{\alpha\beta}K^{\alpha\beta}=R-2\Lambda+\frac{1}{2}\pi_\alpha\pi^\alpha
-\frac{1}{4}F_{\alpha\beta}F^{\alpha\beta}+\frac{1}{2}m^2 A_n^2-\frac{1}{2}m^2 A_\alpha A^\alpha.\,
\end{align}
and
\begin{equation}\label{Simon30}
\nabla_\alpha \pi^\alpha = - m^2 A_n.
\end{equation}
In the above equations the Ricci tensor $R_{\alpha \beta}$ and covariant derivatives $\nabla_\beta$ are those determined by the induced metric $h_{\alpha\beta}$ on a surface of constant $r$. Because we work here in coordinates where the boundary is at $r=0$, the outward-pointing normal one-form is $n = -dr/r$, and consequently there are some sign differences in radial terms relative to \cite{Ross:2011gu}.  $K_{\alpha \beta}$ is the extrinsic curvature of the surface of constant $r$,  $\pi_\alpha = n^\mu F_{\mu\alpha} = -r F_{r\alpha}$ is the conjugate momentum for the massive vector, the radial component of the gauge field is $A_n = n^\mu A_\mu = - r A_r$, and $\dot {}$ denotes the derivative in the normal direction, that is  $-r \partial_r$.

We want to re-express these equations in terms of frame fields $e^A$. As in \cite{Ross:2011gu}, we introduce a frame extrinsic curvature $K^A_{\ B} = e^\alpha_B \dot e^A_\alpha$, which is not a symmetric object, unlike the usual extrinsic curvature. Note that frame indices will be raised and lowered with the metric $g_{AB}$, which is not diagonal in our case, so it is useful to keep track of the `natural' index positions in tensor objects. The equations in frame indices are
\begin{align}
\dot K_{(AB)}+ &
K K_{(AB)} + \frac{1}{2}\left(K_{CA}K^C_{\ B}-K_{AC} K_B^{\ C}\right)
+\frac{1}{2}\pi_A\pi_B-\frac{2}{4(d-2)}\eta_{AB} \pi_C\pi^C
\nonumber\\
&=
R_{AB}-\frac{2}{d-2}\Lambda \eta_{AB}-\frac{1}{2}F_{AC}F_B^{\ C} + \frac{2}{8(d-2)}\eta_{AB} F_{CD}F^{CD}-\frac{1}{2}m^2 A_A A_B,
\\
\dot \pi^A+&
K\pi^A-K^A_{\ B}\pi^B = - \nabla_B F^{BA} + m^2 A^A,
\end{align}
and the constraints
\begin{align}
\nabla^A K_{(AB)}-\nabla_B K^A_{\ A} &=
\frac{1}{2} F_{BA} \pi^A + \frac{1}{2}m^2 A_B A_n,
\\
K^2 -K_{(AB)}K^{AB}-\frac{1}{2}\pi_A \pi^A &=
R-2\Lambda - \frac{1}{4}F_{AB}F^{AB} + \frac{1}{2}m^2 A_n^2 - \frac{1}{2}m^2 A_A A^A, \label{Hconst}
\\
\nabla_A \pi^A &=
 -m^2 A_n.
\end{align}
Here $F_{AB} = e^\alpha_A e^\beta_B F_{\alpha \beta}$, and $\nabla_A = e^\alpha_A \nabla_\alpha$, where the covariant derivative $\nabla_\alpha$  is a total covariant derivative (covariant with respect to both local Lorentz transformations and diffeomorphisms).

Assuming that the metric is asymptotically locally Schrodinger according to the definition \eqref{ALS} then implies that
\begin{equation} \label{Ke}
K^+_{\ \ +} = z + \hat e_+ \dot {\hat e}^+, \quad
K^+_{\ \ -} = r^{2z-2} \hat e_- \dot {\hat e}^+, \quad
K^+_{\ \ I} = r^{z-1} \hat e_I \dot {\hat e}^+,
\end{equation}
\begin{equation} \nonumber
K^-_{\ \ +} = r^{2-2z} \hat e_+ \dot {\hat e}^-, \quad
K^-_{\ \ -} = 2-z + \hat e_- \dot {\hat e}^-, \quad
K^-_{\ \ I} = r^{1-z} \hat e_I \dot {\hat e}^-,
\end{equation}
\begin{equation} \nonumber
K^I_{\ \ +} = r^{1-z} \hat e_+ \dot {\hat e}^I, \quad
K^I_{\ \ -} = r^{z-1} \hat e_- \dot {\hat e}^I, \quad
K^I_{\ \ J} = \delta^I_J + \hat e_J \dot {\hat e}^I.
\end{equation}
Since we choose the frame fields so that the massive vector is $A = \alpha (e^+ + \psi e^- + s_r e^r)$ everywhere in the bulk, the canonical momentum $\pi_A$ has components
\begin{align}
\pi_I = & \alpha (K^+_{\ \ I} +\partial_I s_r),
\nonumber\\
\pi_+ = & \alpha (K^+_{\ \ +} + \partial_+ s_r),
\nonumber\\
\pi_- = & \alpha (\dot \psi +  K^+_{\ \ -} + \partial_- s_r).
\end{align}

To show that a solution exists in an asymptotic expansion, we want to fix the sources, which will fix the terms appearing on the RHS of these equations, and see that we can satisfy the equations by introducing appropriate subleading terms in $r$ in the expansion which will contribute to the radial derivative terms on the LHS of the equations. For this to work, the source terms need to involve positive powers of $r$. Explicit powers of $r$ enter where there are derivatives along the boundary directions: these all come with positive powers of $r$ for $z <2$. There are also explicit powers in the Ricci rotation coefficients, determined by $d e^C = \Omega_{AB}^{\ \ \ C} e^A \wedge e^B$. These are
\begin{equation}
\Omega_{+-}^{\ \ \ +} \sim r^{2-z}, \quad
\Omega_{+I}^{\ \ \ +} \sim r, \quad
\Omega_{-I}^{\ \ \ +} \sim r^{3-2z}, \quad
\Omega_{IJ}^{\ \ \ +} \sim r^{2-z},
\end{equation}
\begin{equation}
\Omega_{+-}^{\ \ \ -} \sim r^{z}, \quad
\Omega_{+I}^{\ \ \ -} \sim r^{2z-1}, \quad
\Omega_{-I}^{\ \ \ -} \sim r, \quad
\Omega_{IJ}^{\ \ \ -} \sim r^{z},
\end{equation}
\begin{equation}
\Omega_{+-}^{\ \ \ I} \sim r, \quad
\Omega_{+J}^{\ \ \ I} \sim r^{z}, \quad
\Omega_{-J}^{\ \ \ I} \sim r^{2-z}, \quad
\Omega_{JK}^{\ \ \ I} \sim r.
\end{equation}
Thus, for $z <2$, the only term that causes problems is $\Omega_{-I}^{\ \ \ +}$, which has a power that becomes negative for $z > 3/2$. This is associated with the $\partial_I$ derivative of the source for $E^\xi$, and the $\partial_-$ derivative of the source for $E^i$. Hence imposing the geometric condition $\hat e^+ \wedge d \hat e^+ = 0$, which will set the sources for $E^\xi$ and $E^i$ to zero, eliminates the leading constribution to this one dangerous term (as well as the leading contribution to $\Omega_{IJ}^{\ \ \ +}$). Note that because of the diffeomorphism invariance, it is only derivatives of these sources that appear. Thus, even though $E^i$ is irrelevant for all $z>1$, the asymptotic expansion exists even in the presence of its source for $1 < z < 3/2$. It is only for $z > 3/2$ that we have to set this source to zero to have a good asymptotic expansion. In addition, a source for the operator dual to the matter field would make a contribution $A_- \sim r^{2-2z}$, so we need to set this source to zero for all $z >1$.

Thus, we expect that an asymptotic expansion will exist for $z < 3/2$ for arbitrary sources in $\hat e^A$ so long as we set  the source for the irrelevant operator $O$ to zero, and for $3/2 < z <2$ if the frame fields satisfy the constraint  $\hat e^+ \wedge d\hat e^+ = 0$ and we set  the source for the irrelevant operator $O$ to zero.

Explicitly analysing the equations of motion is however somewhat messy because of the off-diagonal structure, so we will demonstrate the existence of the asymptotic expansion using the elegant radial Hamiltonian framework of \cite{Papadimitriou:2004ap,Papadimitriou:2004rz}.\footnote{An extended version of this formalism for Lifshitz was introduced in \cite{Chemissany:2014xpa,Chemissany:2014xsa}, but as we work in the frame formalism, we can work simply with an adapted version of the original formalism with a single dilatation operator.}  This involves expanding in eigenvalues of an appropriate bulk dilatation operator. Assuming that we impose some appropriate boundary or regularity condition in the interior of the spacetime, the on-shell solution of the equations of motion will be uniquely determined in terms of the asymptotic boundary data, so the on-shell action is a function of the boundary data, which we can write as a boundary term,
\begin{equation}\label{Simon50}
S=\int d^{d-1}x \sqrt{-\gamma}\lambda(e^{(A)},\psi).
\end{equation}
We can then think of the canonically conjugate momenta as determined by functional derivatives of this action as in a Hamilton-Jacobi approach, so
\begin{equation}\label{Simon52}
T^{A}_{\ \ B} = \frac{1}{\sqrt{-\gamma}}e_\alpha^{(A)}\frac{\delta}{\delta e_{\alpha}^{(B)}}S,
\end{equation}
\begin{equation}\label{Simon53}
\pi_\psi=\frac{1}{\sqrt{-\gamma}}\frac{\delta}{\delta \psi}S.
\end{equation}
For the action \eqref{action}, this gives $T_{AB} = \pi_{AB} + 2 \pi_A A_B$, where $\pi_{AB} = K_{(AB)} - K g_{AB}$. The leading scaling of $\psi$ is $r^{\Delta_-}$, so if we define the dilatation operator
\begin{equation}\label{Simon56}
\delta_D= - \int d^{d_s+2}x \left(
z e_\alpha^{(+)}\frac{\delta}{\delta e_\alpha^{(+)}}+(2-z) e_\alpha^{(-)}\frac{\delta}{\delta e_\alpha^{(-)}}+ e_\alpha^{(I)}\frac{\delta}{\delta e_\alpha^{(I)}}-\Delta_-\psi\frac{\delta}{\delta \psi}
\right).
\end{equation}
then acting on any function of $e^A$, $\psi$, this will agree with the radial derivative at leading order in large $r$, $\delta_D \sim r \partial_r$. Applying this operator to the action, we have
\begin{equation} \label{Simon61}
(d_s+2-\delta_D)\lambda = z T^+_{\ph +}+(2-z)T^-_{\ph -} + T^I_{\ph I}-\Delta_- \psi \pi_\psi.
\end{equation}

Now we look for a solution in an expansion in dilatation eigenvalues $\Delta$. Any function of the boundary data will be by construction an eigenfunction of this dilatation operator, so it will contribute only at one order in the expansion in dilatation eigenvalues. We would then want to expand the action, and hence $T^A_{\ph B}$, $\pi_\psi$, in an expansion in eigenfunctions of the dilatation operator. Because of the coincidences in the powers noted in our linearised analysis, there will be some degenerate eigenvalues, and $\lambda$ does not actually have an expansion in terms of eigenfunctions; the dilatation operator $\delta_D$ is not diagonalisable, but can only be written in a Jordan normal form. This corresponds to the appearance of the logs in the expansion in powers of $r$ in e.g. \eqref{tensor exp}.\footnote{Similar logarithms appear in the Lifshitz case for z=2 \cite{Baggio:2011ha,Zingg:2013xla}; for Schr\"odinger they occur for arbitrary $z$}. However, the first such degenerate eigenvalue occurs at $\Delta = d_s+2$, where the dilatation eigenvalue expansion first makes a finite contribution to the action. Thus, for the purposes of considering the terms that contribute to divergences in the on-shell action, we can expand
\begin{equation}
\lambda = \sum_{d_s +2 > \Delta\geq 0}\lambda^{(\Delta)} + \ldots, \qquad \delta_D\lambda^{(\Delta)}=\Delta \lambda^{(\Delta)}.
\end{equation}
where $\ldots$ represents terms of higher order which will include logarithms.

Let us now set the source for the irrelevant operator $\psi = 0$. Expanding in dilatation eigenvalues, \eqref{Simon61} then becomes
\begin{align}\label{lambdaDeltaeqn}
(d_s+2-\Delta) \lambda^{(\Delta)} = & z T^{+\ph (\Delta)}_{\ph +} + (2-z)T^{-\ph (\Delta)}_{\ph -}+ T^{I\ph (\Delta)}_{\ph I}
\\
= & (4-2z)\pi_{--}^{(\Delta)} + 4\pi_{+-}^{(\Delta)} + 2\pi^{I\ph (\Delta)}_{\ph I}+z  \alpha \pi_-^{(\Delta)} \nonumber
\end{align}

Expanding the constraint equation \eqref{Hconst} in dilatation eigenvalues will enable us to evaluate the RHS of \eqref{lambdaDeltaeqn} in terms of the sources and terms at lower orders in the dilatation expansion. The expansion of \eqref{Hconst} gives
\begin{align}\label{constraintexpanded}
\sum_{s<\Delta/2} &\left[2 K^{(s)}K^{(\Delta-s)}
- 2K_{(AB)}^{(s)}K^{AB(\Delta-s)}-\pi_A^{(s)}\pi^{A(\Delta-s)}-\frac{1}{m^2}(\nabla_A\pi^A)^{(s)}(\nabla_B\pi^B)^{(\Delta-s)}\right]
\\\nonumber
+& \left[K^{(\Delta/2)2}-K_{(AB)}^{(\Delta/2)}K^{AB(\Delta/2)}-\frac{1}{2}\pi_A^{(\Delta/2)}\pi^{A(\Delta/2)}-\frac{1}{2m^2}(\nabla_A\pi^A)^{(\Delta/2)}(\nabla_B\pi^B)^{(\Delta/2)}
\right]=src^{(\Delta)},
\end{align}
where $src^{(\Delta)}$ is the source contribution from the RHS of \eqref{Hconst} which is calculated below in \eqref{src} and following. The terms in the sum at $s=0$, together with one term at $s=\Delta_-$, will give us the RHS of \eqref{lambdaDeltaeqn}. To see this, we need the values of the leading terms in the expansion in dilatation eigenvalues. These are determined by the assumed leading asymptotics of the bulk fields \eqref{ALS}. We have
\begin{equation}
K^{+\ph(0)}_{\ph +}=z, \quad K^{- \ph (0)}_{\ph -}=2-z, \quad K^{I\ph (0)}_{\ph J}=\delta^I_{\ph J}.
\end{equation}
For the vector momentum we have
\begin{equation}
\pi_+^{(0)}= \alpha K^{+\ph (0)}_{\ph +}=z\alpha, \quad \pi_-^{(0)}=0.
\end{equation}
From this we can calculate that
\begin{equation}
T^{A\ph (0)}_{\ph B} = -(d_s +4) \delta^A_{\ph B},
\end{equation}
which is encouraging, as it indicates that this can arise as the functional derivative of a simple constant term, $\lambda^{(0)} = -(d_s+4)$. More importantly, \eqref{lambdaDeltaeqn} can now be combined with \eqref{constraintexpanded} to give
\begin{align}\label{lambdafromsrcandquad}
(d_s+2-\Delta)\lambda^{(\Delta)} = & -src^{(\Delta)}
\\\nonumber
& +\sum_{s<\Delta/2, s \neq 0} \left[-2K_{(AB)}^{(s)}\pi^{AB(\Delta-s)}-\pi_A^{(s)}\pi^{A(\Delta-s)}-\frac{1}{m^2}(\nabla_A\pi^A)^{(s)}(\nabla_B\pi^B)^{(\Delta-s)}\right]
\\\nonumber
& +\left[-K_{(AB)}^{(\Delta/2)}\pi^{AB(\Delta/2)}-\frac{1}{2}\pi_A^{(\Delta/2)}\pi^{A(\Delta/2)}-\frac{1}{2m^2}(\nabla_A\pi^A)^{(\Delta/2)}(\nabla_B\pi^B)^{(\Delta/2)}
\right]
\end{align}

Now let's consider the $src^{(\Delta)}$. We have
\begin{equation} \label{src}
src = R - 2\Lambda -\frac{1}{4}F_{AB}F^{AB}-\frac{m^2}{2}A_A A^A.
\end{equation}
Since we are going to turn $\psi$ off, $A_A A^A=0$, and $F_{AB}$ becomes
\begin{equation}
F_{AB} = 2 \Omega_{AB}^{\ph\ph +} A_+.
\end{equation}
The Ricci scalar is
\begin{equation}\label{Ricciscalar}
R=-4 \partial_A \Omega_{C}^{\ph A C}+\Omega_{CAD}\Omega^{CAD}+2\Omega_{CAD}\Omega^{DAC}+4\Omega_{AD}^{\ph\ph A}\Omega_{C}^{\ph DC},
\end{equation}
which has contributions at $\Delta =2,4-2z,6-4z$, while $F^2$ contributes at just $4-2z$ and $6-4z$. Thus only $-2\Lambda$ contributes to $src^{(0)}$. At $\Delta=2$ we have
\begin{align}
src^{(2)} =
& -4 \partial_+ \Omega_{A-}^{\ph \ph A}-4\partial_- \Omega_{A+}^{\ph \ph A}-4 \partial_I \Omega_{A}^{\ph I A}
+2\Omega_{IJ}^{\ph\ph+}\Omega^{IJ-}+\Omega_{IJK}\Omega^{IJK}
\\\nonumber
& +4\Omega_{+IJ}\Omega_-^{\ph IJ}+4\Omega_{-I}^{\ph \ph +}\Omega_+^{\ph I-}
+ 4 \Omega_{+I}^{\ph \ph +}\Omega_-^{\ph I-}+ 2 \Omega_{+-I}\Omega_{-+}^{\ph \ph I}
+ 4 \Omega_{+-}^{\ph \ph +}\Omega_{-+}^{\ph \ph -}
\\\nonumber
& + 4 \Omega_{A+}^{\ph \ph B} \Omega_{B-}^{\ph \ph A} + 2 \Omega_{AI}^{\ph\ph B}\Omega_B^{\ph IA}+8\Omega_{A+}^{\ph\ph A}\Omega_{B-}^{\ph\ph B}+4\Omega_{AI}^{\ph\ph A} \Omega_B^{\ph IB},
\end{align}
where $A,B$ are taken to run over $+,-$ and all of the $I$ directions.
And for $4-2z$ we find
\begin{align}\label{src4-2z}
src^{(4-2z)} = & -4 \partial_- \Omega_{A-}^{\ph \ph A}
- \Omega_{IJ}^{\ph\ph +}\Omega^{IJ+}-4\Omega_{+I}^{\ph\ph +}\Omega_-^{\ph I+}-2\Omega_{+-}^{\ph\ph +}\Omega_{-+}^{\ph\ph +}
\\\nonumber
&+4\Omega_{-I}^{\ph\ph +}\Omega_-^{\ph I-} + 2\Omega_{-IJ}\Omega_-^{\ph IJ}
+ 2 \Omega_{A-}^{\ph\ph B}\Omega_{B-}^{\ph \ph A}
+4 \Omega_{A-}^{\ph \ph A}\Omega_{B-}^{\ph \ph B}
\\\nonumber
& - \alpha^2 \left(-2(\Omega_{+-}^{\ph\ph +})^2 + 4\Omega_{+I}^{\ph \ph +}\Omega_{-}^{\ph I+}+\Omega_{IJ}^{\ph\ph +}\Omega^{IJ+}
\right).
\end{align}
Lastly for $6-4z$ we have
\begin{align}
src^{(6-4z)} = \left(-2-2 \alpha^2\right)\Omega_{-I}^{\ph\ph +}\Omega_-^{\ph I +}.
\end{align}
For $z <3/2$ this is a positive eigenvalue and we can allow this term, but for $z > 3/2$ it is negative, so we need to restrict the sources so that $\hat e^+ \wedge d \hat e^+ = 0$, so that $src^{(6-4z)} = 0$.

Thus, the source terms will produce contributions to $\lambda^{(\Delta)}$ at $\Delta = 2, 4-2z$ and for $z <3/2$ at $\Delta = 6-4z$. These in turn generate terms in $T^A_{\ph B}$, which we should substitute in the quadratic terms in \eqref{lambdafromsrcandquad} to obtain further contributions to $\lambda$. There are two issues to note here.

The first is that some of the expressions for $K^A_{\ph B}$ in terms of $\hat e^A$ involve explicit positive powers of $r$, so in attempting to solve in a power series in $r$, one might be concerned that having a solution for $K^A_{\ph B}$ in positive powers of $r$ might not necessarily imply that the solution for $\hat e^A$ only involved positive powers of $r$. But by solving first for $\lambda$ and then determining $T^A_{\ph B}$ from it, we avoid this issue. When we functionally differentiate $\lambda$, we pick up a contribution to the dilatation eigenvalue from the different scalings of the different $e^A$, so
\begin{align}\label{Tfromlambda}
\lambda^{(\Delta)} \rightarrow & T^{+\ph (\Delta)}_{\ph +},\quad T^{-\ph (\Delta)}_{\ph -},\quad T^{I\ph (\Delta)}_{\ph J},\quad  T^{+\ph (\Delta+1-z)}_{\ph I}, \quad  T^{-\ph (\Delta+z-1)}_{\ph I},
\\
\nonumber
&  T^{I\ph (\Delta+z-1)}_{\ph +}, \quad  T^{I\ph (\Delta+1-z)}_{\ph -}, \quad  T^{+\ph (\Delta+2-2z)}_{\ph -}, \quad  T^{-\ph (\Delta+2z-2)}_{\ph +}.
\end{align}
The terms where $K^A_{\ph B}$ in terms of $\hat e^A$ involve explicit positive powers of $r$ correspond to those where the functional derivative increases the dilatation eigenvalue. So if we have an expansion in positive powers of $r$ for $\lambda$, it will imply that there is a solution for $\hat e^A$ only involving positive powers of $r$.

Contrariwise, one might be concerned that the functional derivative can also lower the dilatation eigenvalue in \eqref{Tfromlambda}, for $T^+_{\ph -}$, $T^+_{\ph I}$ and $T^I_{\ph -}$. This could lead to contributions to these $T^A_{\ph B}$ with negative dilatation eigenvalues appearing from terms in $\lambda$ with positive dilatation eigenvalues. This could lead to contributions in the sum over quadratic terms in \eqref{lambdafromsrcandquad} with negative eigenvalues, invalidating our assumption that the sum in $\lambda$ involves only positive eigenvalues. For example, differentiating $src^{(6-4z)}$ looks like it could lead to a contribution in $T^+_{\ph -}$ of eigenvalue $8-6z$, which is negative for $z > 4/3$. There is an elegant argument that such a term cannot arise: the stress tensor contribution obtained by this functional derivative is a function of the boundary data, and is a scalar under boundary diffeomorphisms. Any scalar function of the $e^A_\alpha$ can be expressed in terms of the Ricci rotation coefficients $\Omega_{AB}^{\ph \ph C}$, and there is no combination of these coefficients that has this dilatation eigenvalue. Hence the functional derivatives that would give these terms must actually vanish.

It is nice to see this more explicitly however, so we will give the calculation in a couple of cases. We find $T^A_{\ph B}$ from varying with respect to the frame field as in (\ref{Simon52}), with $S$ as in (\ref{Simon50}).  Since the integrand of $S$ contains a factor of $\sqrt{-\gamma}$ we first compute the frame field variation of this term, finding
\begin{equation}
e_\alpha^{(A)}\int \frac{\delta}{\delta e_\alpha^{(B)}} \sqrt{-\gamma} = -\sqrt{-\gamma} \delta^A_{\ph B}.
\end{equation}
Next, since $src$ terms consist of factors of $\Omega_{AB}^{\ph \ph C}$ or $e^\alpha_{(A)}\partial_\alpha$, we work out their variations; functions $f$ are included in these generic expressions to keep track of derivatives in integration by parts. For $\partial_C$ we have
\begin{equation}
e_\alpha^{(A)}\int \frac{\delta}{\delta e_\alpha^{(B)}} f_1 e^\beta_{(C)}\partial_\beta f_2=-f_1 e_\alpha^{(A)}e^\alpha_{(C)}e^\beta_{(B)}\partial_\beta f_2 = -f_1 \delta^A_C\partial_B f_2.
\end{equation}
For variations of $\Omega$ we find
\begin{equation}
e_\gamma^{(D)}\int \frac{\delta \Omega_{AB}^{\ph\ph C}}{\delta e_\gamma^{(E)}} f
=
2\Omega_{E[A}^{\ph\ph C}\delta_{B]}^D f + \delta_E^C\delta_{[A}^D\partial_{B]}f+ \delta_E^C\delta_{[A}^D(\partial_\alpha e^\alpha_{B]})f + f \delta_E^C\Omega_{AB}^{\ph\ph D},
\end{equation}
where $[AB]= \frac{1}{2}(AB-BA)$.

Using these results we can now quickly compute the contribution to $T^A_{\ph B}$ coming from $src^{(6-4z)}$. We find that all of the potentially negative contributions ($T^+_{\ph -}$, $T^+_{\ph I}$ and $T^I_{\ph -}$) actually vanish identically. Considering then the $src^{(4-2z)}$ term, we find this leads to
\begin{equation}
T^{+ (6-4z)}_{\ph -} \propto  -4\left(1+ \alpha^2 \right) \left(\delta \Omega_{+I}^{\ph\ph +}\right)\Omega_-^{\ph I +}+4\Omega_-^{\ph I+}\left(\delta \Omega_{-I}^{\ph\ph -}\right)
\propto  - 4 \alpha^2 \Omega_-^{\ph I+}\Omega_{-I}^{\ph\ph +}.
\end{equation}
As predicted by the general argument, the only possible term is quadratic in $\Omega_{-I}^{\ph\ph +}$.  So if $z>3/2$, where we set this term to zero, no contributions are left.  For $z<3/2$, $T^+_{\ph -}$ does indeed receive this contribution at $\Delta=6-4z$; the contribution is however unproblematic there because it is still at a positive $\Delta$.

The story for $T^+_{\ph I}$ is similar; all terms remaining after the variation have a factor of $\Omega_{-I}^{\ph\ph +}$.  We have
\begin{align}
T^{+ (5-3z)}_{\ph I} \propto & -4 \frac{\partial_-\sqrt{-\gamma}}{\sqrt{-\gamma}}\left( \delta \Omega_{+-}^{\ph\ph +} +\delta \Omega_{J-}^{\ph \ph J}\right)
-4(1+A_+^2) \Omega_-^{\ph J +} \delta \Omega_{+J}^{\ph \ph +}-4(1+A_+^2) \Omega_{+-}^{\ph\ph +} \delta \Omega_{-+}^{\ph\ph +}
\nonumber\\
 & + 4 \Omega_{- \ph J}^{\ph K} \delta \Omega_{-K}^{\ph\ph J}
+ 4 \Omega_{+-}^{\ph\ph +}\delta\Omega_{+-}^{\ph\ph +} + 4 \Omega_{J-}^{\ph \ph +}\delta\Omega_{+-}^{\ph\ph J} + 4 \Omega_{K-}^{\ph\ph J} \delta \Omega_{J-}^{\ph\ph K}
\\\nonumber
& + 8 \Omega_{A-}^{\ph\ph A} \delta \Omega_{+-}^{\ph\ph +} + 8 \Omega_{A-}^{\ph\ph A} \delta \Omega_{J-}^{\ph\ph J}.
\end{align}
Many of the terms here are already multiplied by an $\Omega_{-I}^{\ph\ph +}$.  We need only compute two explicitly:
\begin{align}
\delta\Omega_{+-}^{\ph\ph +}=& -\Omega_{I-}^{\ph\ph+},
\\
\delta\Omega_{K-}^{\ph\ph J} = & \delta_I^J \Omega_{K-}^{\ph\ph+}.
\end{align}
All terms in $T^{+(5-3z)}_I$ coming from $src^{(4-2z)}$ have a factor of $\Omega_{-I}^{\ph\ph +}$.  As in the previous case, for $z>3/2$ this vanishes. For $z<3/2$, $5-3z>0$, and so all of these contributions are at positive $\Delta$ and thus not a concern.

For $T^I_{\ph -}$ we find similarly
\begin{equation}
T^{I (5-3z)}_{\ph -} \propto
-2(1+A_+^2)\Omega^{JK+}\delta\Omega_{JK}^{\ph\ph +}-4(1+A_+^2)\Omega_-^{\ph J+}\delta\Omega_{+J}^{\ph\ph +}
+4\Omega_-^{\ph J+}\delta \Omega_{-J}^{\ph\ph -}.
\end{equation}
Using
\begin{equation}
\delta\Omega_{JK}^{\ph\ph^+} = 2\Omega_{-[J}^{\ph\ph +}\delta^I_{K]},
\end{equation}
we again find that every term in $T^{I (5-3z)}_{\ph -}$ coming from $src^{(4-2z)}$ has a factor of $\Omega_{-I}^{\ph\ph +}$.

Thus, to summarise, there is a solution for $\lambda$ in a series of positive dilatation eigenvalues $\Delta$. Taking functional derivatives of this solution gives the expression for $T^A_{\ B}$ in an expansion in dilatation eigenvalues, which can be used to reconstruct $e^A_\alpha$ in an expansion in positive powers of $r$ which satisfies the equations of motion (with logarithmic terms appearing in the expansion from order $r^{d_s+2}$ onwards, corresponding to the degenerate eigenvalues in the $\Delta$ expansion). The terms in the $\Delta$ expansion of $\lambda$ with $ \Delta < d_s +2$ are the divergent contributions to the bare action, so we also see that we can cancel these terms by adding local functions of the boundary data as boundary counterterms to the action.

\section{Discussion}

We have shown that one can construct a holographic dictionary for $z<2$ Schr\"odinger along very similar lines to the one constructed for Lifshitz in \cite{Ross:2011gu}. This dictionary is based on classifying fields in terms of the anisotropic scaling symmetry of the Schr\"odinger background, unlike some previous explorations of holography for Schr\"odinger which have interpreted it as a deformation of AdS and focused on the relativistic scaling symmetry of the AdS solution. We have shown that in this formalism there is an asymptotic expansion for arbitrary boundary data (assuming we set the sources for irrelevant operators to zero) and the subleading terms in this expansion are all determined locally in terms of the sources.

The most important direction for future work is to extend this analysis to $z=2$, and we aim to address this in a forthcoming paper. As stressed in the introduction, in our frame formalism it is clear that the structure of the dictionary for $z=2$ will be qualitatively different from $z<2$. As already noted in \cite{Guica:2010sw}, the dimensions of operators for $z=2$ depend on the momentum $k_\xi$. We interpret this as meaning that the dual theory will live just in the $t,\vec{x}$ directions, and modes of different $k_\xi$ correspond to different operators in this theory. This will imply a different structure for the dictionary; but we expect the frame formalism will still be useful for organising the bulk modes naturally in terms of the sources for the boundary geometry seen by the field theory, and we expect it will be possible to give an asymptotic expansion at least for arbitrary sources for the $k_\xi=0$ operators.

\section*{Acknowledgements}

We are grateful for helpful conversations with Jelle Hartong, Niels Obers, Blaise Rollier, Kostas Skenderis, Marika Taylor, and Balt van Rees. The work of AP is supported in part by an STFC studentship. The work of TA and SFR is supported by STFC (Consolidated Grant ST/J000426/1). The work of CK is in part supported by the US Department of Energy under grant DE-FG02-95ER40899. CK and SFR thank the Aspen Center for Physics and NSF Grant \#1066293 for hospitality and support during the completion of this work.

\bibliographystyle{JHEP}
\bibliography{Schrodinger}

\providecommand{\href}[2]{#2}\begingroup\raggedright\begin{thebibliography}{10}

\bibitem{Son:2008ye}
D.~Son, {\it {Toward an AdS/cold atoms correspondence: A Geometric realization
  of the Schrodinger symmetry}},  {\em Phys.Rev.} {\bf D78} (2008) 046003,
  [\href{http://xxx.lanl.gov/abs/0804.3972}{{\tt arXiv:0804.3972}}].

\bibitem{Balasubramanian:2008dm}
K.~Balasubramanian and J.~McGreevy, {\it {Gravity duals for non-relativistic
  CFTs}},  {\em Phys.Rev.Lett.} {\bf 101} (2008) 061601,
  [\href{http://xxx.lanl.gov/abs/0804.4053}{{\tt arXiv:0804.4053}}].

\bibitem{Kachru:2008yh}
S.~Kachru, X.~Liu, and M.~Mulligan, {\it {Gravity duals of Lifshitz-like fixed
  points}},  {\em Phys.Rev.} {\bf D78} (2008) 106005,
  [\href{http://xxx.lanl.gov/abs/0808.1725}{{\tt arXiv:0808.1725}}].

\bibitem{Ross:2009ar}
S.~F. Ross and O.~Saremi, {\it {Holographic stress tensor for non-relativistic
  theories}},  {\em JHEP} {\bf 0909} (2009) 009,
  [\href{http://xxx.lanl.gov/abs/0907.1846}{{\tt arXiv:0907.1846}}].

\bibitem{Ross:2011gu}
S.~F. Ross, {\it {Holography for asymptotically locally Lifshitz spacetimes}},
  {\em Class.Quant.Grav.} {\bf 28} (2011) 215019,
  [\href{http://xxx.lanl.gov/abs/1107.4451}{{\tt arXiv:1107.4451}}].

\bibitem{Baggio:2011cp}
M.~Baggio, J.~de~Boer, and K.~Holsheimer, {\it {Hamilton-Jacobi Renormalization
  for Lifshitz Spacetime}},  {\em JHEP} {\bf 1201} (2012) 058,
  [\href{http://xxx.lanl.gov/abs/1107.5562}{{\tt arXiv:1107.5562}}].

\bibitem{Mann:2011hg}
R.~B. Mann and R.~McNees, {\it {Holographic Renormalization for Asymptotically
  Lifshitz Spacetimes}},  {\em JHEP} {\bf 1110} (2011) 129,
  [\href{http://xxx.lanl.gov/abs/1107.5792}{{\tt arXiv:1107.5792}}].

\bibitem{Chemissany:2012du}
W.~Chemissany, D.~Geissbuhler, J.~Hartong, and B.~Rollier, {\it {Holographic
  Renormalization for z=2 Lifshitz Space-Times from AdS}},  {\em
  Class.Quant.Grav.} {\bf 29} (2012) 235017,
  [\href{http://xxx.lanl.gov/abs/1205.5777}{{\tt arXiv:1205.5777}}].

\bibitem{Christensen:2013lma}
M.~H. Christensen, J.~Hartong, N.~A. Obers, and B.~Rollier, {\it {Torsional
  Newton-Cartan Geometry and Lifshitz Holography}},  {\em Phys.Rev.} {\bf D89}
  (2014) 061901, [\href{http://xxx.lanl.gov/abs/1311.4794}{{\tt
  arXiv:1311.4794}}].

\bibitem{Christensen:2013rfa}
M.~H. Christensen, J.~Hartong, N.~A. Obers, and B.~Rollier, {\it {Boundary
  Stress-Energy Tensor and Newton-Cartan Geometry in Lifshitz Holography}},
  {\em JHEP} {\bf 1401} (2014) 057,
  [\href{http://xxx.lanl.gov/abs/1311.6471}{{\tt arXiv:1311.6471}}].

\bibitem{Korovin:2013nha}
Y.~Korovin, K.~Skenderis, and M.~Taylor, {\it {Lifshitz from AdS at finite
  temperature and top down models}},  {\em JHEP} {\bf 1311} (2013) 127,
  [\href{http://xxx.lanl.gov/abs/1306.3344}{{\tt arXiv:1306.3344}}].

\bibitem{Nishida:2007pj}
Y.~Nishida and D.~T. Son, {\it {Nonrelativistic conformal field theories}},
  {\em Phys.Rev.} {\bf D76} (2007) 086004,
  [\href{http://xxx.lanl.gov/abs/0706.3746}{{\tt arXiv:0706.3746}}].

\bibitem{Guica:2010sw}
M.~Guica, K.~Skenderis, M.~Taylor, and B.~C. van Rees, {\it {Holography for
  Schrodinger backgrounds}},  {\em JHEP} {\bf 1102} (2011) 056,
  [\href{http://xxx.lanl.gov/abs/1008.1991}{{\tt arXiv:1008.1991}}].

\bibitem{Costa:2010cn}
R.~Caldeira~Costa and M.~Taylor, {\it {Holography for chiral scale-invariant
  models}},  {\em JHEP} {\bf 1102} (2011) 082,
  [\href{http://xxx.lanl.gov/abs/1010.4800}{{\tt arXiv:1010.4800}}].

\bibitem{Guica:2011ia}
M.~Guica, {\it {A Fefferman-Graham-Like Expansion for Null Warped AdS(3)}},
  \href{http://xxx.lanl.gov/abs/1111.6978}{{\tt arXiv:1111.6978}}.

\bibitem{vanRees:2012cw}
B.~C. van Rees, {\it {Correlation functions for Schrodinger backgrounds}},
  \href{http://xxx.lanl.gov/abs/1206.6507}{{\tt arXiv:1206.6507}}.

\bibitem{Hartong:2013cba}
J.~Hartong and B.~Rollier, {\it {Particle Number and 3D Schroedinger
  Holography}},  \href{http://xxx.lanl.gov/abs/1305.3653}{{\tt
  arXiv:1305.3653}}.

\bibitem{Rangamani:2008gi}
M.~Rangamani, S.~F. Ross, D.~Son, and E.~G. Thompson, {\it {Conformal
  non-relativistic hydrodynamics from gravity}},  {\em JHEP} {\bf 0901} (2009)
  075, [\href{http://xxx.lanl.gov/abs/0811.2049}{{\tt arXiv:0811.2049}}].

\bibitem{Papadimitriou:2010as}
I.~Papadimitriou, {\it {Holographic renormalization as a canonical
  transformation}},  {\em JHEP} {\bf 1011} (2010) 014,
  [\href{http://xxx.lanl.gov/abs/1007.4592}{{\tt arXiv:1007.4592}}].

\bibitem{Andrade:2013wsa}
T.~Andrade and S.~F. Ross, {\it {Boundary conditions for metric fluctuations in
  Lifshitz}},  {\em Class.Quant.Grav.} {\bf 30} (2013) 195017,
  [\href{http://xxx.lanl.gov/abs/1305.3539}{{\tt arXiv:1305.3539}}].

\bibitem{Papadimitriou:2004ap}
I.~Papadimitriou and K.~Skenderis, {\it {AdS / CFT correspondence and
  geometry}},  \href{http://xxx.lanl.gov/abs/hep-th/0404176}{{\tt
  hep-th/0404176}}.

\bibitem{Papadimitriou:2004rz}
I.~Papadimitriou and K.~Skenderis, {\it {Correlation functions in holographic
  RG flows}},  {\em JHEP} {\bf 0410} (2004) 075,
  [\href{http://xxx.lanl.gov/abs/hep-th/0407071}{{\tt hep-th/0407071}}].

\bibitem{Chemissany:2014xpa}
W.~Chemissany and I.~Papadimitriou, {\it {Generalized dilatation operator
  method for non-relativistic holography}},
  \href{http://xxx.lanl.gov/abs/1405.3965}{{\tt arXiv:1405.3965}}.

\bibitem{Chemissany:2014xsa}
W.~Chemissany and I.~Papadimitriou, {\it {Lifshitz holography: The whole
  shebang}},  \href{http://xxx.lanl.gov/abs/1408.0795}{{\tt arXiv:1408.0795}}.

\bibitem{Baggio:2011ha}
M.~Baggio, J.~de~Boer, and K.~Holsheimer, {\it {Anomalous Breaking of
  Anisotropic Scaling Symmetry in the Quantum Lifshitz Model}},  {\em JHEP}
  {\bf 1207} (2012) 099, [\href{http://xxx.lanl.gov/abs/1112.6416}{{\tt
  arXiv:1112.6416}}].

\bibitem{Zingg:2013xla}
T.~Zingg, {\it {Logarithmic two-point correlation functions from a z =2
  Lifshitz model}},  {\em JHEP} {\bf 1401} (2014) 108,
  [\href{http://xxx.lanl.gov/abs/1310.4778}{{\tt arXiv:1310.4778}}].

\end{thebibliography}\endgroup

\end{document}